# On Compensating Tune Spread Induced by Space Charge in Bunched Beams

Vladimir N. Litvinenko*[1,2] and Gang Wang[2]

[1] Department of Physics and Astronomy, Stony Brook University, Stony Brook, NY 11794, USA

[2] Collider Accelerator Department, Brookhaven National Laboratory, Upton, NY 11973, USA

* Corresponding author: vladimir.litvinenko@stonybrook.edu

**Abstract.**

Space charge effects play significant role in modern-day accelerators. These effects frequently constrain attainable beam parameters in an accelerator - or - in an accelerator chain. They also could limit the luminosity of hadron colliders operating either at low energies or with a sub-TeV high brightness hadron beams. The later is applied for strongly cooled proton and ion beams in eRHIC – the proposed future electron-ion collider at Brookhaven National Laboratory [1].

A number of schemes for compensating space charge effects in a coasting (e.g. continuous) hadron beam were proposed and some of them had been tested. Using a proper transverse profile of the electron beam (or plasma column) for a coasting beam would compensate both the tune shift and the tune spread in the hadron beam. But all of these methods do not address the issue of tune spread compensation of a bunched hadron beam, e.g. the tune shift dependence on the longitudinal position inside the bunch.

In this paper we propose and evaluate a novel idea of using a co-propagating electron bunch with miss-matched longitudinal velocity to compensate the space charge induced tune-shift and tune spread. We present a number of practical examples of such system.

## I.    Introduction.

Space charge effects are known in accelerator physics for a half of the century. There is extensive literature [2-22] (with an excellent and concise review by Zotter in [23]) describing space charge affect on the beam's quality and stability. Nonlinear space-charge force induces an irreducible transverse tune spread, e.g. the tune dependence on both the hadron's[1] longitudinal position inside the bunch, $z$, and the amplitude of the transverse oscillations.

It is well known that space charge effects fall a high power of the beam's relativistic factor:

---

[1] Here we are considering only a positively charge particles whose space charge effects can be compensated using negatively charged electrons. The case of negatively charged particles, including antiprotons and negatively charged ion, would require positively charged particles for such compensation. Using position, proton or ion beams for compensating space charge effects in the negatively charged beams, while theoretically possible, is, most likely, impractical.

$$\Delta Q_{sc} \approx -\frac{Z^2 r_p}{A} \frac{N_o}{4\pi \beta_h^2 \gamma^3 \varepsilon} \frac{C}{\sqrt{2\pi}\sigma_z} \tag{1}$$

where $C$ is the ring circumference, $Z$ is the charge and $A$ is the atomic number of the hadron (e.g. an ion, for proton $Z = A = 1$), $r_p = e^2/m_p c^2$ is the classical radius of the proton, $\gamma^2 = 1/(1-\beta^2)$ is the relativistic factor of hadron beam, $N_o$ is number of hadrons in the bunch with RMS duration of $\sigma_z$, and $\varepsilon$ it the transverse emittance of the beam.

While space charge effects exist in any charged beam, they have stronger implications for hadron beams. Hadrons become ultra-relativistic at much higher energies and also travel longer pass in accelerators, compared with their lepton counterpart, to become ultra-relativistic.

One of the most important effects is the tune spread induced by intrinsically nonlinear space charge force[2]. General expression of the tune spread is given in Appendix A. Since the compensation scheme we are presenting here does not depend on the details of the transverse beam distribution, e.g. similarly to compensation techniques suggested for coasting beams, the transverse profile of the electron beam (or a column) should compensate both the tune shift and its dependence on the amplitudes of the transverse oscillations [24-28]. Naturally the maximum tune shift is experienced by the particles in the center of the beam, while the particles with large amplitude of oscillations have a smaller value of the tune shift. The overall tune spread is determined by its value for the center particles.

Thus, for simplicity we will consider here a hadron beam with equal transverse emittances $\varepsilon_x = \varepsilon_y = \varepsilon$, whose tune shifts are given by (A20):

$$\delta Q_{x,y} = \delta Q_{sc}(z) \cdot f_{x,y};$$

$$\delta Q_{sc}(z) = -\frac{C}{4\pi\varepsilon} \frac{1}{\beta^2 \gamma^3} \frac{Z^2 r_p}{A} \cdot \frac{N_o}{\sqrt{2\pi}\sigma_z} \cdot e^{-\frac{z^2}{2\sigma_z^2}}; \tag{2}$$

$$f_x = \left\langle \frac{2}{1+\sqrt{\beta_y/\beta_x}} \right\rangle; f_y = \left\langle \frac{2}{1+\sqrt{\beta_x/\beta_y}} \right\rangle.$$

Since longitudinal motion of hadrons is usually very slow (e.g. $Q_s \ll Q_{x,y}$), the tune of the particle depends not only on the amplitudes (actions) of the transverse oscillations, but also on longitudinal the location within the bunch.

One practically important feature of the space charge effects is a very strong dependence on the relativistic factor $\gamma$: $\delta Q_{sc} \propto \gamma^{-3}/(1-\gamma^{-2})$. While the power one of the $\gamma$ naturally comes from increasing rigidity of the beam, the $\gamma^{-2}$ comes from effective canceling of the forces from electric and magnetic fields induced by the beam (details are in Appendix A):

---

[2] In this paper, for compactness, we assume a Gaussian longitudinal distribution of particles. Naturally, the treatment presented in this paper can be extended to other types of longitudinal distributions.

$$\vec{F}_\perp = eZ\left(\vec{E}_\perp + \beta_o\left[\hat{z}\times\vec{B}_\perp\right]\right) = eZ\cdot\vec{E}_\perp\left(1-\beta_o^2\right) \equiv \frac{eZ\cdot\vec{E}_\perp}{\gamma^2}. \qquad (3)$$

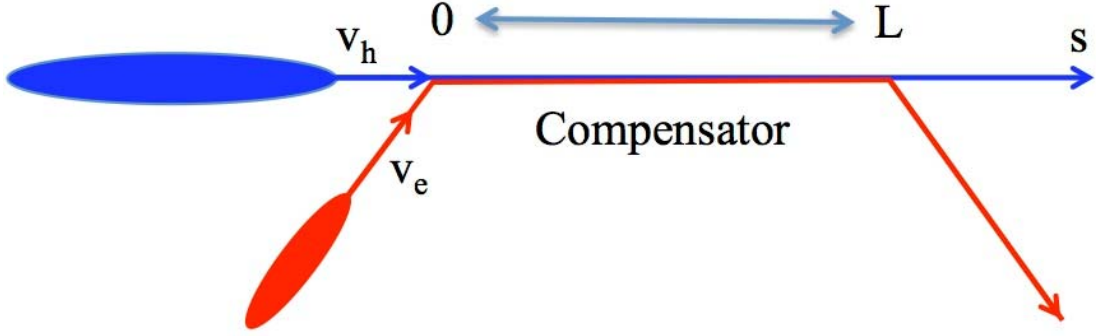

Fig. 1. A generic layout of the space charge compensator interaction region. Electron beam merges with the hadron beam, co-propagates through straight section, and then extracted.

A number of practical schemes for space-charge tune shift and tune spread compensation using an electron beam colliding with hadron beam (e.g. an electron lens) or an electron column induced in a residual gas [24-28] were suggested. The tune shift given by the colliding beam does not suffer from $\gamma_h^{-2}$ cancelation. Wise versa - it is amplified. For an electron lens the interaction length of $L$, the tune shift is:

$$\delta Q_{x,yel} = \frac{Z}{A}\frac{1}{\beta_h\gamma_h}\frac{r_p}{4\pi\sigma_e^2}\cdot\frac{I_e}{ec\beta_e}\left(1+\beta_e^2\right)\cdot L\langle\beta_{x.y}\rangle;$$

$$L\langle\beta_{x.y}\rangle \equiv \int_0^L \beta_{x.y}\,dz; \qquad (4)$$

where $\beta_e = v_e/c$ is the normalized velocity of electron beam[3]. Comparing eqs. (2) and (4) one can conclude that electron beam current

$$I_e = \frac{C}{2\gamma_h^2 L}\cdot\frac{2\beta_e}{\beta_h\left(1+\beta_e^2\right)}\cdot\frac{\sigma_e^2}{\langle\beta_{x.y}\rangle\varepsilon_h}\cdot I_p \propto \frac{C}{2\gamma_h^2 L}\cdot\frac{\beta_e}{\beta_h}\cdot I_p$$

$$I_p = \frac{ecZN_o}{\sqrt{2\pi}\sigma_z};\ \frac{\sigma_e^2}{\langle\beta_{x.y}\rangle\varepsilon_h}\sim 1. \qquad (5)$$

can be used to compensate for the space charge tune shift. Typically the interaction length is significantly smaller than the ring circumference (to be exact, it is always smaller), e.g. $\eta = L/C \ll 1$. This shortcoming can be compensated by large relativistic factors $2\gamma_h^2 \beta_h / \beta_e \gg 1$. It means that the electron current in such electron lens can be modest, and frequently, comparable with the hadron beam current.

---

[3] We would like to point to a possible confusion caused by a multiple usage of symbol $\beta$. Unfortunately it is unavoidable when one has to use both the velocities $\beta = v/c$ and the lattice functions $\beta_{x,y}$ in the same paper.

As explained in [24,28], selecting a proper transverse distribution of electron beam, one can match the space charge tune shift dependence on the transverse amplitudes. The only, but important, short-coming of this method is that the tune-shift introduced by the electron lens (or the column) is identical for all particles independent on the longitudinal location inside the bunch.

In a bunched beam, however, the space-charge tune shift depends on the hadron's position within the hadron bunch, $z$. Thus, the $z$-dependence of the tune shift cannot be compensated using an electron lens or an electron column. Thus, for a bunched beam, these schemes could, at best, reduce the space-charge tune spread by a half, e.g. by compensating it at the half of the peak value at $z=0$.

Using a co-propagating electron beam with the same relativistic velocity $\beta_e = \beta_h$ (as in electron cooling schemes) and the same longitudinal distribution offers an opportunity of compensating both the transverse and longitudinal dependence of the space charge field. Unfortunately, the compensating beam suffers from $\gamma_h^{-2}$ cancelation, and such scheme would require a very large electron beam current:

$$I_e \approx \frac{C}{L} \cdot I_p \gg I_p. \tag{6}$$

This unfavorable scaling makes such a scheme impractical, especially for hadron beams in large colliders. For example, eRHIC would operating hadron beams with peak current ~ 10 A (and duration of 0.4 nsec). Using a 30 m of 3.8 km RHIC circumference for such space charge compensator would require an electron bunch with peak current ~ 1.2 kA and the bunch charge ~ 4,000 nC. Such e-beam simply does not exist.

We propose to use the co-propagating scheme, but with mismatched relativistic factors of the two beams. Such mode provides for a possibility to diminish the reduction factor while keeping the slippage between the beams under control.

## II. The idea of the method.

The idea of the proposed method is based on a simple observation that the relativistic canceling is proportional to $\gamma^2$, while the velocity of particles weakly depends of $\gamma$ for $\gamma > 2$. To be exact, we consider a co-propagating relativistic e-beam having nearly identical bunch profile as the hadrons but having a different relativistic factor.

Let's consider first a simple case of both beams being round and relativistic, e.g. both beam velocities are close to the speed of light. Hence, the slippage of the e-beam with the respect to the hadron beam is rather small compared with the length of the interaction section, $L$:

$$\Delta z = \left(v_e - v_h\right)\tau = \frac{L}{\beta_h}\left(\beta_e - \beta_h\right) \cong \frac{L}{2}\left(\frac{1}{\gamma_h^2} - \frac{1}{\gamma_e^2}\right) \tag{7}$$

For long bunches, the fields from the both beams can be easily calculated using the Gauss law:

$$B_{h\theta}(r,z,s) = \frac{2I_o(z)}{cr}\int_0^r f_h(x,s)xdx; \quad E_{hr}(r,z,s) = -\frac{B_{h\theta}(r,z,s)}{\beta_h};$$

$$B_{e\theta}(r,z,s) = \frac{2I_e(z)}{cr}\int_0^r f_e(x,s)xdx; \quad E_{er}(r,z,s) = -\frac{B_{e\theta}(r,z,s)}{\beta_e};$$

where $f_{e,h}(r,s)$ are transverse distributions of the beams. The force acting on the hadron from the e-beam is:

$$\begin{aligned} F_r &= Ze\left(-\frac{B_{e\theta}(r,z,s)}{\beta_e} + \beta_h B_{e\theta}(r,z,s)\right) = -\frac{Ze}{\beta_e}B_{e\theta}(r,z,s)\cdot(1-\beta_e\beta_h) \\ &= -Ze\frac{2I_e(z)}{cr}\int_0^r f_e(x,s)xdx \cdot \frac{1-\sqrt{(1-\gamma_e^{-2})(1-\gamma_h^{-2})}}{\sqrt{(1-\gamma_e^{-2})}} \end{aligned} \quad (8)$$

For ultra-relativistic particles (8) becomes:

$$F_r \cong -Ze\frac{I_e(z)}{cr}\int_0^r f_e(x,s)xdx \cdot \left(\frac{1}{\gamma_e^2}+\frac{1}{\gamma_h^2}\right) \quad (9)$$

while self-action gives:

$$F_{r\,sc} \cong Ze\frac{I_h(z)}{cr}\int_0^r f_h(x,s)xdx \cdot \frac{2}{\gamma_h^2}. \quad (10)$$

Thus, in the case of unequal velocities, the relativistic $\gamma_h^{-2}$ cancelation is replaces by $(\gamma_e^{-2}+\gamma_h^{-2})/2$.

Since using low energy electron beams is economically favorable, let's assume that relativistic factor of hadron is significantly larger than that of the e-beam:

$$\gamma_h^2 \gg \gamma_e^2.$$

This assumption makes $\frac{1}{\gamma_e^2} \gg \frac{1}{\gamma_h^2}$ and simplifies (7) and (9):

$$\begin{aligned} F_{hr} &\cong -Ze\frac{2I_h(z)}{\gamma_h^2 cr}\int_0^r f_h(x,s)xdx; \\ F_{er} &\cong -Ze\frac{I_e(z)}{\gamma_e^2 cr}\int_0^r f_e(x,s)xdx; \\ \Delta z &\cong -\frac{L}{2\gamma_e^2}. \end{aligned} \quad (11)$$

Let's assume the e-beam has the same transverse shape as the hadron beam. Then, as follows from eq. (11), compensating for the space charge effects accumulated by the hadron beam in the ring with circumference C, we will need the electron beam current of interacting with the hadrons in a straight section[4] with length L to be:

$$I_e \cong -2I_h \cdot \frac{\gamma_e^2}{\gamma_h^2} \cdot \frac{C}{L} \qquad (12)$$

Let's further assume that it could slip for one RMS bunch-length of the proton bunch (see practical examples in following Section) during the interaction. e.g. the length of interaction section is:

$$L \sim 2\gamma_e^2 \sigma_z. \qquad (13)$$

Then, the ratio between the beam currents becomes independent of the electron's energy:

$$\frac{I_e}{I_h} \cong \frac{C}{\sigma_z \gamma_h^2}, \qquad (14)$$

and the later should be determined by a practical matters. Naturally, there can be $N_i \geq 1$ space charge compensation sections[5], which proportionally reduce the required e-beam current.

$$\frac{I_e}{I_h} \cong \frac{C}{N_i \sigma_z \gamma_h^2}, \qquad (14')$$

For example, $\gamma_h = 100$ and RHIC circumference $C$=3.8 km the required ratio is:

$$\frac{I_e}{I_h} \sim \frac{1}{N_i} \cdot \frac{1}{\sigma_z(n\sec)} \qquad (15)$$

It means that electron beam peak current can of the same order at that of the hadrons.

### III. The method.

Let's now consider the method for finite velocities without any limitations. Let's consider the electron beam having longitudinal profile determined by it current

$$I_e\left(t - \frac{s}{v_e}\right). \qquad (16)$$

It merges with the hadron beam, co-propagates along the interaction region from s=0 and taken out at s=L. A hadron passes the interaction region as follows:

---

[4] It is unlikely that the same magnets can be used to equally bend trajectories of electrons and hadrons when $\gamma_h > \gamma_e$.

[5] As we discuss later, it actually can be even beneficial.

$$s = v_h \cdot (t - t_o); \quad t = t_o + \frac{s}{v_h}; \tag{17}$$

and is affected by the electron beam current of

$$I_e \left( t_o + \frac{s}{v_h} - \frac{s}{v_e} \right). \tag{18}$$

The integrated effect is expressed by the following expression:

$$L \cdot \overline{I}_e(t) = \int_o^L I_e \left( t + \frac{s}{v_h} - \frac{s}{v_e} \right) ds \equiv c \frac{\beta_e \beta_h}{\beta_h - \beta_e} \int_o^{\Delta t} I_e(t + \zeta - \Delta t) d\zeta \tag{19}$$

with the slippage given by

$$c\Delta t = L \frac{\beta_h - \beta_e}{\beta_e \beta_h}. \tag{20}$$

First, let's assess the value of allowable slippage by deconvolving equation (19) assuming that the shape of $\overline{I}_e(t)$ repeats that to the hadron beam $I_h(t)$, e.g.:

$$\frac{\overline{I}_e(t)}{\overline{I}_e(0)} = \frac{I_h(t)}{I_h(0)} = q(t) \tag{21}$$

Thus we will require that

$$\int_0^{\Delta t} I_e(t - \Delta t + \zeta) d\zeta = I_o \cdot q(t) \tag{21}$$

with value of $I_o$ to be chosen to compensate the tune shift for the hadron in the center of the bunch. Any deviation of e-bunch the shape from (21) will result in error of compensating tune for the hadrons. In Appendix D we show how to deconvolving

$$\int_0^{\Delta t} g(t + \zeta) d\zeta = q(t); \quad I_e(t - \Delta t) = I_o g(t); \tag{22}$$

to get a simple two independent solutions:

$$g_+(t) = -\sum_{m=0}^{\infty} q'(t + m\Delta t);$$
$$g_-(t) = +\sum_{m=1}^{\infty} q'(z - m\Delta t). \tag{23}$$

It is obvious observation a linear combination of the solutions (23)

$$g_\alpha(t) = \alpha g_+(t) + (1 - \alpha) g_-(t)$$

is a solution of (22). It is likely that $g_{1/2} = (g_+ + g_-)/2$ can be of practical interest.

We also shown in Appendix D that for a rather general physics assumptions these functions converge to zero at one of the infinities:

$$g_+(z)_{z \to +\infty} \to 0; \quad g_-(z)_{z \to -\infty} \to 0 \qquad (24)$$

It is not necessarily true about the other sign.

While these mathematical properties of the solutions are mostly of academic interest, there is an additional, and very practical, issue. By definition in eq. (21) $q(t)$ is non-negative function: $q(t) \geq 0$. Similarly, the sign of the e-beam current and $I_e(t) \leq 0$. Thus, any practical Deconvolution of (22) cannot change the sign, and choosing $I_o < 0$ requires $g(t)$ being a positive function:

$$g(t) \geq 0. \qquad (25)$$

The natural parameter determining behavior and "positivity" of $g_\pm(t)$ is determined by the ratio between the slippage, $\Delta t$, and the RMS length of the hadron bunch, $\sigma_t$,:

$$\tau = \frac{\Delta t}{\sigma_t}. \qquad (26)$$

In storage rings longitudinal distribution is frequently described by a Gaussian function:

$$q(t) = \exp\left[-\frac{t^2}{2\sigma_t^2}\right]. \qquad (27)$$

We study this distribution in detail (see Appendix D) to answer following questions:

(a) at what values of $\tau$ deconvolution functions remain positively defined;

(b) what error is accumulated in the convolution if we fit a reasonable positively defined function to approximate $g_\pm(t)$ for large values of $\tau$.

The following is a short summary of out findings. First, for $\tau \leq 1$, both $g_\pm(t)$ solutions behave and converge very well within the typical physical aperture (in our case we used ± 5 RMS bunch lengths). Deconvolutions $g_\pm(t)$ are nearly identical (see Fig. 2) and positively defined in the interval $t/\sigma_t \in \{-5, 5\}$ For $\tau = 0.5$, the difference between $g_\pm(t)$ is within $\pm 10^{-15} g_\pm(0)$ and is likely determined by the computer accuracy. For $\tau = 1$, the difference between $g_\pm(t)$ is less than $\pm 10^{-7} g_\pm(0)$ and $g_\pm(5\sigma_t) \cong 1.8 \cdot 10^{-5}$, $g_\pm(-5\sigma_t) \cong 9.5 \cdot 10^{-8}$. It simply means that for practical purposes the compensating error will be defined not by accuracy the deconvolution function, but by other practical means.

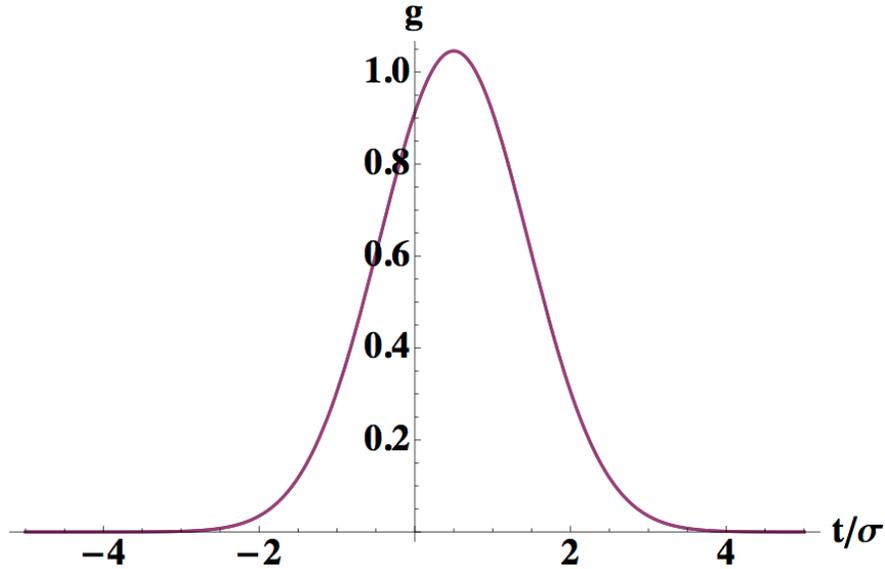

Fig. 2. Graphs of $g_\pm(t)$ for deconvolution for Gaussian distribution (27) with $\tau = 1$. One should notice that both functions $g_\pm(t)$ are practically indistinguishable.

Second, for values of $\tau$ exceeding unity, the situation changes rather rapidly and when for $\tau > 1.75$ a very well defined oscillating tails with amplitude comparable with the central peak. Naturally both $g_\pm(t)$ are no longer positively defined. Fig. 3 and 4 visualize these features.

(a)  (b)

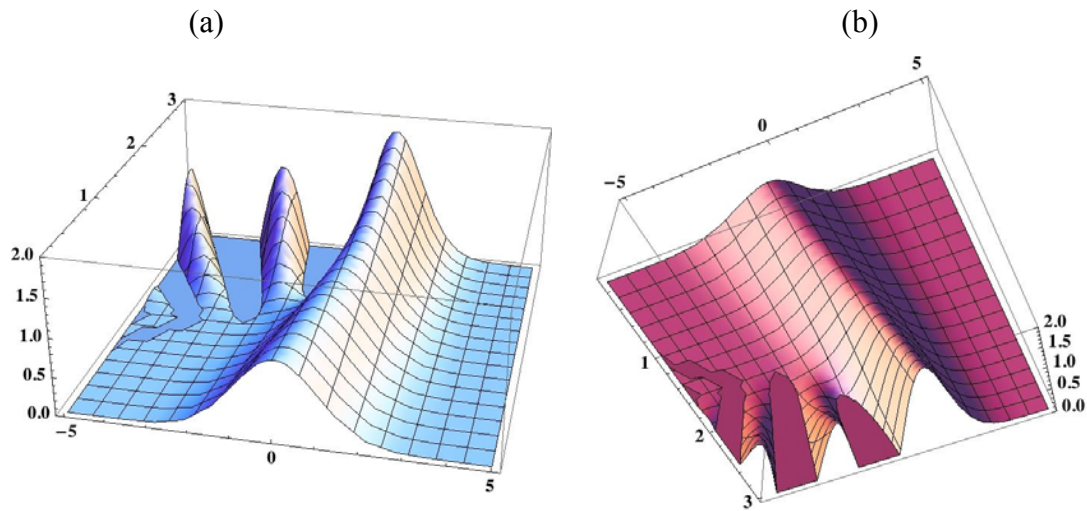

Fig. 3. 3D graphs of deconvolution function $g_+$ for Gaussian distribution (27) at function of the time and slippage. (a) is the top view with horizontal axis being $t/\sigma_t$ {-5.5}, the vertical axis being $g_+(t)$, and the third axis used for the slippage $\tau = \Delta t / \sigma_t$. The vertical axis is clipped at zero to clearly indicate where $g_+$ becomes negative. (b) the same graph seen from the bottom to clearly indicate the areas of $g_+ < 0$.

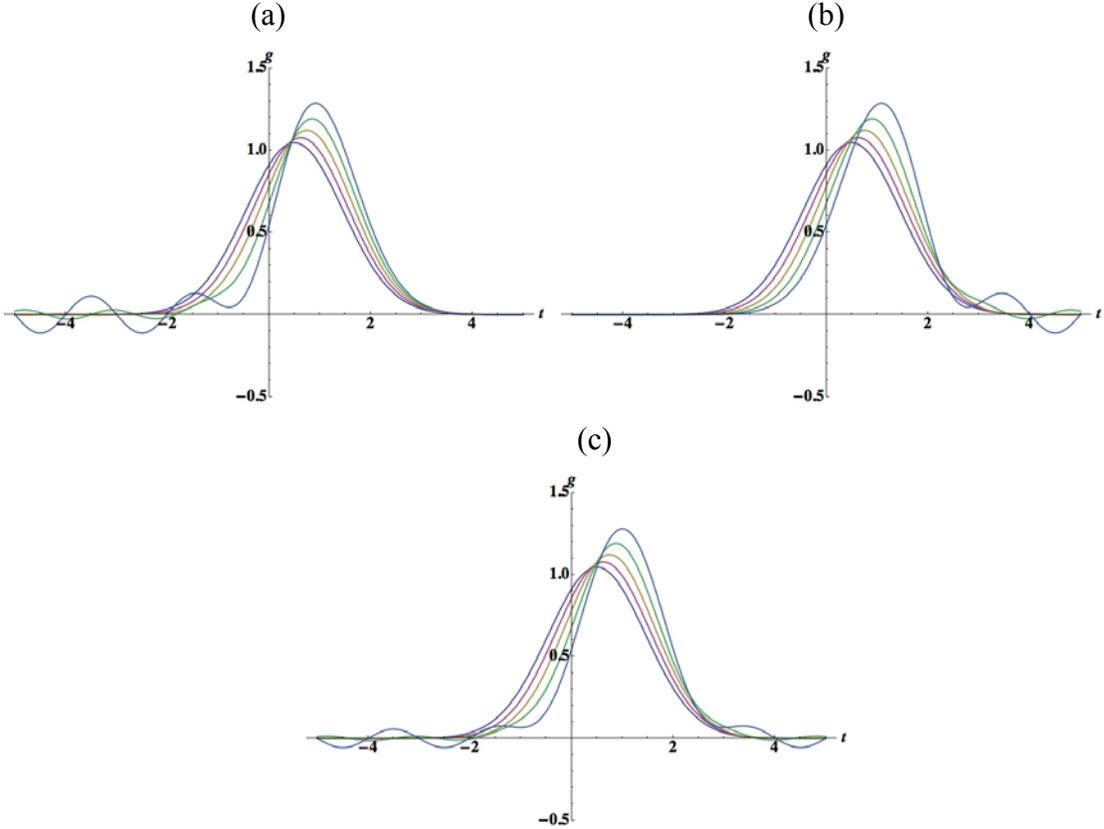

Fig. 4. 3D graphs of $g_+$ (a), $g_-$ (b) and $g_{1/2}$ (c) deconvolution functions for a Gaussian distribution (27) at function of the time for four values of $\tau = \Delta t / \sigma_t$: 1 – dark blue, 1.25 – magenta, 1.5 – yellow/grey, 1.75 – green and 2 – light blue.

Practical conclusion form this studies is that $\tau = 1.5$ is a natural boundary, where a $g_{1/2}$ deconvolution (see Fig.5) working very well. It provides a relative convolution error of less than $10^{-3}$ (to be exact its is limited to about $5 \times 10^{-4}$) and, therefore, would not represent practical accuracy limit. For example, a bunch to bunch variation in a hadron beam intensity most likely will exceed relative level of $10^{-3}$.

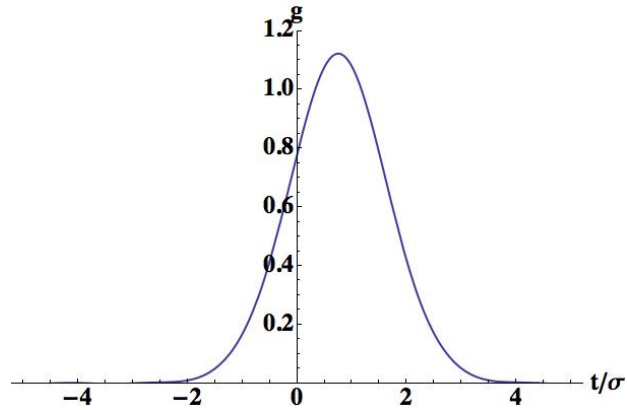

Fig. 5. Clipped graph of $g_{1/2}$ for a case of $\tau = 1.5$. Keeping the shape of $g_{1/2}$ from $t = -3\sigma_t$ to $t = 4\sigma_t$ provides for a nearly perfect convolution (22).

As shown in Appendix D, even for $\tau = 2$ we could find a positive function (by fitting $g$), which could compensate 95% of the tune spread in the beam. Nevertheless, for the rest of the paper we will use $\tau = 1.5$ as a practical limit for the slippage.

Since, we proven that for a smooth Gaussian-line longitudinal distribution of the hadron bunch we can nearly perfectly compensate the space-charge induced tune spread, the question is what electron beam current is required to do this?

Issues related to the transverse matching of the nonlinear tune shift induced by the space charge in the ring and that induced by the electron beam in the compensator are discussed in Appendices A, B and C. Here, for simplicity, let's further consider round beam space charge effects and set $f_{x,y} = 1$ in eq. (2)[6]. In this case, the space charge tune shift accumulated by hadrons along the ring circumference C should be compensated by that accumulated in the interaction with electrons:

$$\delta Q_{sc} = -\frac{C}{4\pi\varepsilon}\frac{1}{\beta_h^3\gamma_h^3}\frac{Zr_p}{A}\cdot\frac{I_h}{ec} = -\delta Q_{sce} = \frac{L_c}{8\pi\varepsilon}\frac{1}{\beta_h\gamma_h}\frac{1-\beta_e\beta_h}{\beta_e}\frac{Zr_p}{A}\cdot\frac{\overline{I_e}}{ec} \qquad (28)$$

This, the requirement on the compensating beam current:

$$\frac{L_c}{\beta_e}\overline{I_e}\cdot(1-\beta_e\beta_h) = I_h C\frac{(1-\beta_h^2)}{\beta_h} \rightarrow \frac{\overline{I_e}}{I_h} = -\frac{C}{L_c}\frac{\beta_e}{\beta_h}\frac{(1-\beta_h^2)}{(1-\beta_e\beta_h)} \qquad (29)$$

depends on the length of the compensator section, $L_c$. The length $L_c$ could be limited either by the allowable slippage, as we discussed above, $c\Delta t \leq 1.5\sigma_z/\beta_h$:

$$L_{\max} \leq c\Delta t_{\max}\frac{\beta_e\beta_h}{\beta_h-\beta_e} \approx \frac{1.5\sigma_z\beta_e}{\beta_h-\beta_e}; \qquad (30)$$

or by practical limitations of the accelerator. Combined with the limitation on the slippage gives:

$$\left|\frac{\overline{I_e}}{I_h}\right| = \frac{C}{\gamma_h^2\beta_h^2(1-\beta_e\beta_h)}\max\left\{\frac{|\beta_e-\beta_h|}{\Delta tc},\frac{\beta_e\beta_h}{L_c}\right\}. \qquad (31)$$

As we discussed in previous sections, there is no benefits of having $\beta_e \geq \beta_h$. Hence, for $\beta_e < \beta_h$ eq. (23) becomes:

$$\left|\frac{\overline{I_e}}{I_h}\right| = \frac{C}{\gamma_h^2\beta_h^2}\max\left\{\frac{\beta_h-\beta_e}{c\Delta t(1-\beta_e\beta_h)},\frac{\beta_e\beta_h}{L_c(1-\beta_e\beta_h)}\right\}. \qquad (32)$$

---

[6] Compensation criteria do not depend on the details of the transverse matching. The later is required to compensate correctly space charge tune spreads in horizontal and vertical directions, as well as to approximate the space charge tune shift dependences on the transverse actions $I_{x,y}$.

To minimize the required electron beam current we can find minimum of the right-hand-side in eq. (32). Let's us note, that $\dfrac{\beta_h - \beta_e}{1 - \beta_e \beta_h} = \beta_h \dfrac{1 - \beta_e/\beta_h}{1 - \beta_e \beta_h}$ monotonically reduces and $\dfrac{\beta_h \beta_e}{1 - \beta_e \beta_h}$ monotonically increases as function of $\beta_e$ at the interval $0 \leq \beta_e \leq \beta_h$. Thus the minimum in eq. (32) is reached at

$$\beta_e = \beta_c = \dfrac{\beta_h}{1 + \beta_h \dfrac{c\Delta t}{L_c}}; \qquad (33)$$

This yields a simple expression for the required compensating current:

$$\dfrac{\overline{I}_e}{I_h} = -\dfrac{C}{L_c + \gamma_h^2 \beta_h \cdot c\Delta t}. \qquad (34)$$

In a case that there is more than one compensator $N_c \geq 1$, the required e-beam current can be reduced proportionally:

$$\dfrac{\overline{I}_e}{I_h} \cong \dfrac{1}{N_c} \dfrac{C}{L_c + \gamma_h^2 \beta_h \cdot c\Delta t} \qquad (35)$$

Having more than one compensator may have additional advantages: it will distribute compensation around the ring. The later will bring compensators closer to the source and naturally will provide more stable beam (see discussions at the end of the paper).

According to eq. (35) using multiple compensators with the given total length of

$$L = N_c L_c = \eta C \qquad (36)$$

there required e-beam would be:

$$\dfrac{\overline{I}_e}{I_h} \cong \dfrac{1}{\eta + N_c \cdot \gamma_h^2 \beta_h \cdot \dfrac{c\Delta t}{C}}. \qquad (37)$$

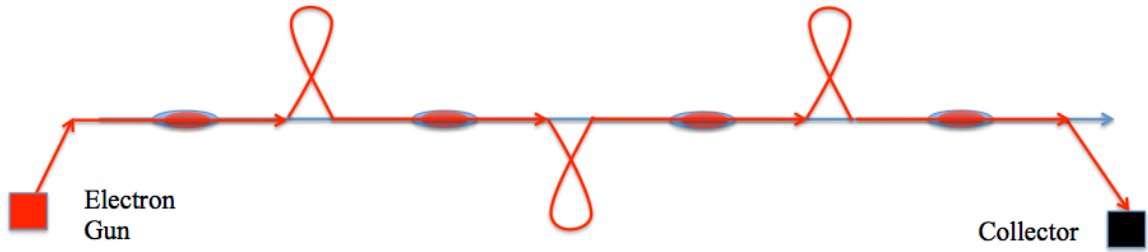

Fig. 6. A sketch of one electron source supporting four SC compensating sections. Each electron bunch (red) merges and co-propagate with the hadron beam in the first straight section. Then it sent through a delay loop to merge with the following hadron bunch. The process then repeated as needed (three time in this figure).

Thus, it is beneficial to split compensation length into as many as practically manageable and possible compensators. As follows from (33) it will also lower the optimal electron beam energy.

Such splitting will not necessarily lead to the increase of the electron beam sources and collector. Hadron storage rings usually operate with many hadrons bunches. For example RHIC (and future eRHIC) operate with 120 hadron bunches. Fig. 6 shows a scheme in which one electron beam source can serve a number of space charge compensators.

This scheme is especially effective when a low energy electron beam is used moving with non-relativistic velocities. In following section we will discuss what can set a limit on the minimal length of compensator.

### IV.   Effects on the electron beam

It is well known that hadron beam will affect the propagation and dynamic of the electron beam. There is a natural desire, driven by economy, of using low-energy electron beams. Such beam would be highly susceptible to space charge forces induced by itself as well as by the hadron beam. The solution for transporting such beam and accurate control of their distribution is well known: the magnetized electron beam transported by a solenoidal field [29-40]. Such transport provides both the stability of the electron (and therefore, the hadron) beam as well as control of the beam size by changing the value of the solenoidal field. The solenoidal field provides the focusing to counteract destructive space charge and collective interaction.

For example, the stability of the interacting electron and hadron beam will be similar to that in head-on collisions using electron lens [42-45,40]. As shown in [46], using strong longitudinal magnetic field plays important role in maintaining stability of interacting electron and hadron beams.

If solenoidal transport is used, the main limiting factor for the compensator length will be a finite radius of curvature of the bends. At each bend, the hadron experiences the field from the bent electron beam, which has a different transverse structure from that that we are compensating. Thus, the effect of the end-effects has to be controlled and we can formulate an addition limitation on the compensator length:

$$L_c \gg bend\,radius$$

In practice it meant that typical lengths of a compensator should be few meter and not few centimeters.

### V.   Examples.

Let's consider a case a hadron beam in eRHIC with $\gamma_h = 250$, an RMS bunch length of $\sigma_z$ =8 cm and the compensator length of 3 m. We assume that that slippage is equal to 1.5 RMS bunch length, e.g. $c\Delta t = 12 cm$. Equation (33) yields the optimum $\beta_e \cong 0.961$ and the optimum electron beam energy of 1.86 MeV (kinetic energy of 1.35 MeV). The required electron beam current for a single compensator should be:

$$\overline{I}_e \cong -0.51 \cdot I_h \quad (36)$$

In high-luminosity eRHIC the expected proton intensity is $3 \cdot 10^{11}$ per bunch and the peak proton beam current of ~72 A. Hence, the required e-beam peak current ~ *37* A is reasonable. For a $\gamma_h = 50$ the required e-beam current would grow nearly 12.6-fold to about *900* A, making such compensation in a single compensator a very challenging task.

Table 1. Examples of possible space-charge compensator schemes for various mode of operating the High-Luminosity eRHIC

| Parameter | | | | |
|---|---|---|---|---|
| **Hadron beam** | p | p | $^{79}Au^{197}$ | $^{79}Au^{197}$ |
| Energy, GeV/u | 250 | 100 | 100 | 50 |
| Number of particles | $3\ 10^{11}$ | $3\ 10^{11}$ | $3\ 10^9$ | $3\ 10^9$ |
| RMS bunch length, m | 0.08 | 0.08 | 0.08 | 0.08 |
| Norm. emittance, m rad | $2\ 10^{-7}$ | $2\ 10^{-7}$ | $2\ 10^{-7}$ | $2\ 10^{-7}$ |
| $\Delta Q_{sc}$ | 0.05 | 0.31 | 0.1 | 0.40 |
| Peak current, A | 72 | 72 | 58 | 58 |
| **Electron beam** | | | | |
| Compensator length, m | 3 | 3 | 3 | 3 |
| Kinetic Energy, MeV | 1.35 | 1.35 | 1.35 | 1.35 |
| # of compensators | 3 | 12 | 12 | 24 |
| Peak current, A | 10.6 | 16.6 | 13.1 | 26 |
| Total comp. length, m | 9 | 36 | 36 | 74 |

Hence the length of the compensation section is only 3 m, while each of six eRHIC IR straight sections is 200 m long, one can consider multiple compensating sections to reduce the required e-beam peak current to tens of amps.

Compensating space charge effects could also be required for low-energy scan in RHIC in search of critical point in the QCD phase-diagram [47-51]. For such scan the relativistic γ can be as low as 2.7 [52], but bunch intensity in such operation is significantly lower than in eRHIC and the bunch length is much longer [53]. Tables below give a sample of hadron beam energies, hadron beam parameters and possible parameters for the space-charge compensation.

Parameters of the compensators presented in the tables are not necessarily optimized for the performance or cost. What is important that energy of electron beam required for space charge compensation for RHIC's low energy scan is in ~10 KeV range (this was the reason to quote kinetic energy in the table). This is directly related to use of a long bunches with RMS duration of few meters. Long bunches allow a longer slippage, and therefore operating e-beams with $\beta_e \sim 0.2 - 0.25$. This also means electrons are rather slow and that loops in Fig. 6 are 4-5 fold shorter than would be required for relativistic electrons.

Table 2. Examples of possible space-charge compensator schemes for various mode of low energy RHIC operations *

| Parameter | | | | |
|---|---|---|---|---|
| **Hadron beam** | $^{79}Au^{197}$ | $^{79}Au^{197}$ | $^{79}Au^{197}$ | $^{79}Au^{197}$ |
| Energy, GeV/u | 2.5 | 3.85 | 5.75 | 10 |
| Number of particles | $0.5 \, 10^9$ | $1.1 \, 10^9$ | $1.1 \, 10^9$ | $1.1 \, 10^9$ |
| RMS bunch length, m | 3 | 3 | 2 | 2 |
| Norm. emittance, m rad | $1.5 \, 10^{-6}$ | $1.5 \, 10^{-6}$ | $1.5 \, 10^{-6}$ | $1.5 \, 10^{-6}$ |
| $\Delta Q_{sc}$ | 0.10 | 0.09 | 0.06 | 0.075 |
| Peak current, A | 0.25 | 0.55 | 0.82 | 0.83 |
| **Electron beam** | | | | |
| Compensator length, m | 1 | 1 | 1 | 1 |
| Kinetic Energy, keV | 8.6 | 8.6 | 16.5 | 16.5 |
| # of compensators | 12 | 12 | 12 | 12 |
| Peak current, A | 2.5 | 2.4 | 2.4 | 0.77 |
| Total comp. length, m | 12 | 12 | 12 | 12 |

*Ion beam parameters are taken from [52-53] and we assume that the beam transverse emittance is cooled by a factor of 2 [54-55].*

In contrast, eRHIC will use short bunches of hadrons with RMS length of few centimeters. It results is a modest but relativistic energy of electron ~ 1.35 MeV with $\beta_e \cong 0.96$.

These simple examples show the capability of this concept to compensate of the space charge induced tune spread with reasonable length of the compensator and reasonable current of the compensator.

## VI. Discussion and Conclusions

In this paper we presented a novel method of compensating space-charge-induced tune spread in bunched hadron beams. We showed that, in principle, it is possible to compensate both the tune shift and the tune spread with a significant accuracy. We consider that with a proper design and simulations, up to 99% of the space-charge-induced tune spread can be compensated.

It is a natural to ask a question of what would be an ultimate of the space-charge-induced tune spread, which can be sustained in a storage ring? Unfortunately there is no simple direct answer. It is well know that space-charge-induced resonances can make hadron beam unstable [56-58]. Thus, the beam dynamics in the presence of the compensators should be simulated using appropriate code (see , for example [59-63]).

One important conclusion of such simulations (for the bunch average tune shift compensation [59-62]) is that spreading the compensators around the ring allow to achieve (at least in simulation) larger beam intensities. It is a frequently observed phenomenon that local compensation for nonlinear effects in beam dynamics, e.g. placing the compensation as close as possible to the source, is preferable solution.

Hence, we believe that having multiple tune spread compensators spread around the ring would result in better compensation and beam stability. For example, one can chose the strength and the locations of the compensators to eliminate most dangerous resonant driving terms.

In contrast with traditional space-charge compensating schemes, we proposed the method of compensating not only the tune shift but also the entire tune spread. We expect that this method would allow stabilizing the intense hadron beam in sub-TeV and low energy storage rings. Electron beams with required quality, energy and intensity either are readily available or considered for future accelerators [64-67] and could be used for the proposed space-charge effects compensator.

## VII.    Acknowledgements

Authors are indebted to Alexey Fedotov (BNL) for indispensible help with directing us toward the most recent studies and publications of space-charge related effects. We also appreciate in-depth discussion of relevant formulae. We are grateful to Ilan Ben-Zvi and Thomas Roser (both BNL) for their interest in our approach and thoughtful comments and suggestions. We would like to thank Vladimir Shiltsev (FNAL) for providing us with latest copy of his presentation on various space-charge compensation ideas, such as electron lens and plasma column.

Our special thanks belong to Alexander Pikin, who shared with us his deep knowledge of the beam parameters and shapes attainable with low energy electron beams.

**Appendix A. Space charge induced tune spread**

There is a multiple ways of deriving the space-charge tune shift and tune spread for relativistic hadron beam. We refer interested reader to specialized papers on the topic [68-69] and references therein. There is no practical limit to which one can complicate the space charge problem by adding effects of the surrounding environment [70], non-trivial beam distributions, and of cause, an arbitrary coupling between three degrees of freedom. While interesting in general, heavy mathematical calculations could obscure the main idea of this paper, e.g. the describing the novel compensation method for the space charged induced tune spread.

Hence, we will focus on a case of uncoupled transverse motion with as simple Gaussian transverse and longitudinal density distribution in the bunch propagating in vacuum:

$$f(x,y,s,\tau=ct) = \frac{N_o}{(2\pi)^{3/2} \sigma_x \sigma_y \sigma_z} \exp\left(-\frac{x^2}{2\sigma_x^2} - \frac{y^2}{2\sigma_y^2} - \frac{(s-\beta_o\tau)^2}{2\sigma_z^2}\right); \qquad (A1)$$

where $N_o$ is the number of particles in the bunch, $\beta_o \equiv (1-\gamma_o^{-2})^{1/2} = v_o/c$ is the beam longitudinal velocity, $\tau = ct$ and $c$ is the speed of the light. Trivial Lorentz transformation gives us the distribution (A1) in the co-moving frame:

$$\bar{f}(x,y,\bar{z}) = \frac{N_o}{(2\pi)^{3/2} \gamma_o \sigma_x \sigma_y \sigma_z} \exp\left(-\frac{x^2}{2\sigma_x^2} - \frac{y^2}{2\sigma_y^2} - \frac{\bar{z}^2}{2\bar{\sigma}_z^2}\right); \qquad (A2)$$

$$\bar{z} = \gamma_o(s-\beta_o\tau); \quad \bar{\sigma}_z = \gamma_o \sigma_z.$$

The charge density $\rho$ differs from the particle density by the multiplier $eZ$. It is our assumption that in the co-moving frame the scalar potential is nearly time-impendent, e.g. it evolves only with the change of the particles distribution. Naturally, since there is no current in the co-moving frame, the vector potential from space charge is equal zero. Now we just need to solve stationary Poisson equation [7]:

$$\Delta \bar{\varphi}(\vec{r}) = -4\pi \rho(\vec{r}).$$

Following [71] we can derive the scalar potential using well known equalities [72-73]:

$$\bar{\varphi}(\vec{r}) = \int \frac{\rho(\vec{\zeta})}{|\vec{r}-\vec{\zeta}|} d\vec{\zeta}^3; \quad \frac{1}{|\vec{r}-\vec{\zeta}|} \equiv \frac{2}{\sqrt{\pi}} \int_0^\infty \exp\left(-u^2 |\vec{r}-\vec{\zeta}|^2\right) du$$

and rewrite it using $q=1/u^2$ as:

$$\bar{\varphi}(\vec{r}) = \frac{1}{\sqrt{\pi}} \int_0^\infty \int q^{-3/2} \rho(\vec{\zeta}) \exp\left(-\frac{|\vec{r}-\vec{\zeta}|^2}{q}\right) dq d\vec{\zeta}^3.$$

---

[7] We assume here that in the co-moving frame speed of the particles is much smaller that the speed of the light and the formula for a static scalar potential $\Delta\varphi = -4\pi\rho$ is applicable.

Substituting the charge density using the particle distributions (A2):

$$\bar{\varphi}(x,y,\bar{z}) = \frac{eZ}{\sqrt{\pi}} \frac{N_o}{(2\pi)^{3/2} \gamma_o \sigma_x \sigma_y \sigma_z}$$

$$\int_0^\infty \frac{dq}{q^{3/2}} \int d\xi d\eta d\zeta \cdot \exp\left(-\frac{(x-\xi)^2 + (y-\eta)^2 + (\bar{z}-\zeta)^2}{q} - \frac{\xi^2}{2\sigma_x^2} - \frac{\eta^2}{2\sigma_y^2} - \frac{\zeta^2}{2\bar{\sigma}_z^2}\right); \quad \text{(A3)}$$

and taking three trivial integrals, as indicated below:

$$\frac{(x-\xi)^2}{q} + \frac{\xi^2}{2\sigma_x^2} = \frac{q+2\sigma_x^2}{2\sigma_x^2 q}\left(\xi - x\frac{2\sigma_x^2}{q+2\sigma_x^2}\right)^2 + \frac{x^2}{q+2\sigma_x^2};$$

$$\int_{-\infty}^{\infty} d\xi \exp\left(-\frac{(x-\xi)^2}{q} - \frac{\xi^2}{2\sigma_x^2}\right) = \sigma_x \sqrt{\frac{2\pi q}{q+2\sigma_x^2}} \exp\left(-\frac{x^2}{q+2\sigma_x^2}\right);$$

we arrive to final expression for a scalar potential:

$$\bar{\varphi}(x,y,\bar{z}) = \varphi_o + \frac{eZ}{\sqrt{\pi}} N_o \cdot \int_0^\infty dq \frac{\exp\left(-\frac{x^2}{q+2\sigma_x^2} - \frac{y^2}{q+2\sigma_y^2} - \frac{\bar{z}^2}{q+2\bar{\sigma}_z^2}\right)}{\sqrt{(q+2\sigma_x^2)(q+2\sigma_y^2)(q+2\bar{\sigma}_z^2)}}. \quad \text{(A4)}$$

Since in the co-moving frame only time-component of the 4-vector potential, e.g. $\bar{\varphi}$, is non-zero, the Lorenz transformation back into the lab-frame is trivial:

$$\left(\varphi_{sc}(x,y,s,\tau), \vec{A}_{sc}(x,y,s,\tau)\right)_{lab} = \gamma_o(1, \beta_o \hat{s}) \cdot \bar{\varphi}(x,y,\gamma(s-\beta_o \tau)), \quad \text{(A5)}$$

where $\hat{s}$ is the unit vector along the beam orbit. Eq. (A5) can be directly added into the Canonical accelerator Hamiltonian [74-75]:

$$h^* = -(1+Kx)\sqrt{\frac{(H-e\varphi+e\varphi_{sc})^2}{c^2} - m^2c^2 - \left(P_x - \frac{e}{c}A_x\right)^2 - \left(P_y - \frac{e}{c}A_y\right)^2} - \frac{e}{c}(A_z + A_{cs}) \quad \text{(A6)}$$

where we assumed a flat ring reference orbit (i.e. absence of the orbit torsion [74]).

While it is possible to proceed a bit further using the Hamiltonian (A6), for most practical cases we can use significant simplifications. First, in all practical hadron storage rings the bunch length is significantly larger compared with the transverse beam sizes, e.g. $\sigma_z \gg \sigma_{x,y}$. For example in eRHIC the hadron beam RMS bunch length will be ~ 50 mm, while the transverse beam size will be ~ 0.2 mm. With further $\bar{\sigma}_z = \gamma_o \sigma_z$ boost in the co-moving frame this size asymmetry becomes overwhelming, e.g.

$$\bar{\sigma}_z \ggg \sigma_{x,y}.$$

In eRHIC case we would have $\bar{\sigma}_z / \sigma_{x,y} > 10^4$. This asymmetry allows using a two-dimensional expression for the scalar potential. The easiest way is using $\bar{\sigma}_z^2 \to \infty$ limit in (A4), but it can be also done directly [76] [8]:

$$\bar{\varphi}(x,y,\bar{z}) = \frac{eZN_o}{\sqrt{2\pi}\bar{\sigma}_z} e^{-\frac{\bar{z}^2}{2\bar{\sigma}_z^2}} \cdot \int_0^\infty dq \frac{\exp\left(-\frac{x^2}{q+2\sigma_x^2} - \frac{y^2}{q+2\sigma_y^2}\right)}{\sqrt{(q+2\sigma_x^2)(q+2\sigma_y^2)}}. \quad (A7)$$

Using substitutions $u(s) = \sigma_x(s)/\sigma_y(s)$ and $q = 2\vartheta\sigma_x\sigma_y$. Eq. (A7) can be rewritten with a dimensionless integral:

$$\bar{\varphi}(x,y,\bar{z}) = \frac{eZN_o}{\sqrt{2\pi}\bar{\sigma}_z} e^{-\frac{\bar{z}^2}{2\bar{\sigma}_z^2}} \cdot \int_0^\infty d\vartheta \frac{\exp\left(-\frac{x^2/\sigma_x^2}{2(1+\vartheta/u)} - \frac{y^2/\sigma_y^2}{2(1+\vartheta u)}\right)}{\sqrt{(1+\vartheta/u)(1+\vartheta u)}}; \quad (A8)$$

The near-the-axis expansion of eq. (A7) can be found written in analytical form using an identity $a(a+b)\int_0^\infty dq / \sqrt{(q+2a^2)^3 (q+2b^2)} = 1$:

---

[8] There is a direct way of doing deriving (A7) following [76]. By applying a Fourier transform $\iint .. \exp(\vec{k}\vec{r}) dxdy/(2\pi)^2$ to $\Delta_\perp \varphi \approx -4\pi\rho$; for $\bar{\sigma}_z \gg \sigma_{x,y}$ we attain $\varphi(\vec{k}) = 4\pi\rho(\vec{k})/\vec{k}^2$.

Using familiar trick $\frac{1}{\vec{k}^2} = \int_0^\infty e^{-\vec{k}^2 t} dt \equiv \frac{1}{4}\int_0^\infty e^{-\frac{\vec{k}^2 t}{4}} dt$ and scaling it by 1/4$^{th}$ one gets:

$$\varphi(\vec{k}) = 4\pi \int_0^\infty \rho(\vec{k}) e^{-\vec{k}^2 t} dt; \quad \varphi(\vec{r}) = \pi \int_0^\infty dt \iint e^{-i\vec{k}\vec{r}} \rho(\vec{k}) e^{-\frac{\vec{k}^2 t}{4}} dk_x dk_y.$$

Then for a long bunch Gaussian bunch with linear density of $\rho_o(z) = eZN \cdot e^{-\frac{\bar{z}^2}{2\bar{\sigma}_z^2}} / (\sqrt{2\pi}\bar{\sigma}_z)$.

$$\rho = \rho_o(z) \frac{1}{2\pi\sigma_x\sigma_y} e^{-\frac{x^2}{2\sigma_x^2} - \frac{y^2}{2\sigma_y^2}}; \quad \rho(k) = \frac{1}{(2\pi)^2} \cdot e^{-\frac{k_x^2\sigma_x^2}{2} - \frac{k_y^2\sigma_y^2}{2}};$$

after trivial integration

$$\varphi(\vec{r}) = \pi\rho_o(z) \frac{1}{(2\pi)^2} \cdot \int_0^\infty dt \iint e^{-i\vec{k}\vec{r}} e^{-\frac{k_x^2\sigma_x^2}{2} - \frac{k_y^2\sigma_y^2}{2}} e^{-\frac{\vec{k}^2 t}{4}} dk_x dk_y; \quad \int_{-\infty}^\infty e^{-ik_x x} e^{-\frac{k_x^2(2\sigma_x^2+t)}{4}} dx = \sqrt{\frac{4\pi}{2\sigma_x^2+t}} e^{-\frac{x^2}{2\sigma_x^2+t}};$$

we get the desirable result identical to (A7):

$$\varphi(\vec{r}) = \frac{eZN}{\sqrt{2\pi}\bar{\sigma}_z} e^{-\frac{z^2}{2\sigma_z^2}} \cdot \int_0^\infty \frac{e^{-\frac{x^2}{2\sigma_x^2+t} - \frac{y^2}{2\sigma_y^2+t}}}{\sqrt{(2\sigma_x^2+t)(2\sigma_y^2+t)}} dt.$$

$$\bar{\varphi}(x,y,\bar{z}) = \bar{\varphi}_o(z) - \frac{eZN_o}{\sqrt{2\pi}\bar{\sigma}_z} e^{-\frac{\bar{z}^2}{2\bar{\sigma}_z^2}} \cdot \frac{1}{\sigma_x + \sigma_y}\left(\frac{x^2}{\sigma_x} + \frac{y^2}{\sigma_y}\right) + O^4$$

$$\bar{\varphi}_o(z) = \frac{eZN_o}{\sqrt{2\pi}\bar{\sigma}_z} e^{-\frac{\bar{z}^2}{2\bar{\sigma}_z^2}} \cdot \int_0^\infty \frac{dq}{\sqrt{(q+2\sigma_x^2)(q+2\sigma_y^2)}}$$

(A9)

Next useful approximation, which very common in accelerator literature originates from the fact that synchrotron oscillations in hadron rings are much slower compared with the transverse (betatron) oscillation. It allows on to consider both the longitudinal position of particle inside the bunch as well it relative energy deviation, $\delta \equiv E/E_o - 1$, as slow varying. In this approximation the transverse accelerator Hamiltonian (with dimensionless variables $p_{x,y} \to p_{x,y}/p_o$) becomes [74, 75]:

$$\tilde{h} = -(1+Kx)\sqrt{1 + \frac{2}{\beta_o}\left(\delta - \frac{Ze}{p_o c}\varphi_{sc}\right) + \left(\delta - \frac{Ze}{p_o c}\varphi_{sc}\right)^2 - p_x^2 - p_y^2} - \frac{Ze}{pc}(A_s + A_{sc}); \quad (A10)$$

$$\left(\varphi_{sc}(x,y,s,\tau), \vec{A}_{sc}(x,y,s,\tau)\right)_{lab} = \gamma_o(1, +\beta_o \hat{s}) \cdot \bar{\varphi}(x, y, \gamma(s - \beta_o \tau));$$

where $K$ is the curvature of the reference orbit.

Assuming that space charge can be treated locally as a perturbation (e.g. $\left|\Delta Q_{sc\,x,y}\right| \ll Q_{x,y}$), we can expand the Hamiltonian keeping only dominant space charge term. Using (A5) and taking into account that $\gamma_o(1 - \beta_0^2) = \gamma_o^{-1}$ we get:

$$\tilde{h} = \tilde{h}_o + \Delta\tilde{h}_{sc} \cong -(1+Kx)\sqrt{1 + \frac{2\delta}{\beta_o} + \delta^2 - \vec{p}_\perp^2} - \frac{Ze}{pc}A_s + \left\{\frac{1+Kx}{1+\delta/\beta_o}\right\} \cdot \frac{Ze}{\beta_o \gamma_o p_o c}\bar{\varphi}(x, y, \gamma(s - \beta_o \tau)).$$

(A11)

Since we are interested in the main space charge effects, keeping the term in angular brackets is unnecessary[9].

---

[9] The multiplier $1/(1+\delta/\beta_o)$ describing the chromaticity of the space charge effects and can be estimated as $\sigma_\delta \cdot \Delta Q_{sc} \sim 10^{-3}\Delta Q_{sc}$, while the multiplier $1+Kx$ represents the lengthening of the trajectory and can be estimated as $\alpha_c \sigma_\delta \cdot \Delta Q_{sc} \sim 10^{-3}\Delta Q_{sc}/\gamma_t^2$, where $\gamma_t$ is the relativistic factor at transition energy. Typically $\gamma_t \sim Q_x \gg 1$. More accurate treatment requires averaging perturbation of the Hamiltonian over the phases of the betatron oscillations and integrating it over the ring circumference, $C$. It gives us an effective one-turn variation of average Hamiltonian [75]:

$$\Delta\tilde{h}_{sc} = \frac{1+Kx}{\beta_o + \delta} \cdot \frac{e}{\gamma_o p_o c}\bar{\varphi}(x,y,\bar{z}); \quad h_{cs}(I_x, I_y, z, \delta) = \frac{e}{\gamma_o p_o c}\frac{1}{\beta_o + \delta}\int_0^C ds \langle(1+Kx) \cdot \bar{\varphi}(x,y,\bar{z})\rangle_{\phi_{x,y}};$$

$$x = \sqrt{2\beta_x(s)I_x}\cos(\psi_x(s) + \phi_x) + D_x\delta; \quad y = \sqrt{2\beta_y(s)I_y}\cos(\psi_y(s) + \phi_y)$$

where $I_x, I_y$ are the actions of the betatron oscillations, $\delta$ is the relative energy deviation of the hadron and $\beta_{x,y}$ are the lattice-functions. The action dependent tune shift than can be calculated as:

In (A11) the third term represents space charge effects:

$$\Delta\tilde{h}_{sc} = \frac{1}{\beta_o^2\gamma_o^3}\frac{Z^2 N_o r_c}{\sqrt{2\pi}\sigma_z}e^{-\frac{(s-\beta_o\tau)^2}{2\sigma_z^2}}\cdot\int_0^\infty d\vartheta\frac{\exp\left(-\frac{x^2/\sigma_x^2}{2(1+\vartheta/u)}-\frac{y^2/\sigma_y^2}{2(1+\vartheta u)}\right)}{\sqrt{(1+\vartheta/u)(1+\vartheta u)}}; \quad (A12)$$

where $r_c = e^2/mc^2$ is the classical particle's radius. Since we are using $Ze$ as the charge the hadrons, we should also introduce the hadron's mass number $A = m/m_p$ (with $A > 1$ for ions), where $m_p$ is the mass of the proton: $r_c = r_p/A$; $r_p = e^2/m_p c^2$.

Near the axis expansion of eq. (A12) is straightforward using (A9):

$$\Delta\tilde{h}_{scL} = -\frac{1}{\beta_o^2\gamma_o^3}\frac{Z^2 N_o r_p}{A\sqrt{2\pi}\sigma_z}e^{-\frac{(s-\beta_o\tau)^2}{2\sigma_z^2}}\left(\frac{x^2}{\sigma_x(\sigma_x+\sigma_y)}+\frac{y^2}{\sigma_y(\sigma_x+\sigma_y)}\right). \quad (A13)$$

The classical averaging method[10] over the fast betatron oscillations [74, 75]

$$\Delta h_{sc}(I_x, I_y) = \langle\Delta\tilde{h}_{sc}(x,y)\rangle_{\phi_x,\phi_y};$$
$$x = \sqrt{2\beta_x(s)I_x}\cos(\psi_x(s)+\phi_x);\ y = \sqrt{2\beta_y(s)I_y}\cos(\psi_y(s)+\phi_y). \quad (A14)$$

where $I_{x,y}$ and $\phi_{x,y}$ are the actions and phases of the betatron oscillations, allows to find local variation of the betatron phases as derivatives over the corresponding action:

$$\frac{d\phi_{x,y}}{ds} = \frac{\partial\Delta h_{sc}(I_x, I_y, s)}{\partial I_{x,y}}.$$

Then the tune shifts is a simple integral over the ring circumference:

$$\Delta Q_{sc\,x,y} = \frac{1}{2\pi}\oint_o ds\frac{\partial\Delta h_{sc}(I_x, I_y, s)}{\partial I_{x,y}}. \quad (A15)$$

---

$$\Delta Q_{sc\,x,y} = \frac{1}{2\pi}\frac{\partial h_{cs}(I_x, I_y, z, \delta)}{\partial I_{x,y}}.$$

This formula covers all aspects of the space-charge tune shift, including its chromaticity. For on energy particles $\delta = 0$:

$$h_{cs}(I_x, I_y, z) = \int_0^\oint ds\left\langle\frac{1+Kx}{\gamma_o\beta_o}\cdot\frac{e}{p_o c}\bar\varphi(x,y,\bar z)\right\rangle_{\phi_{x,y}};$$
$$x = \sqrt{2\beta_x(s)I_x}\cos(\psi_x(s)+\phi_x);\ y = \sqrt{2\beta_y(s)I_y}\cos(\psi_y(s)+\phi_y).$$

Since the potential is symmetric functions $\varphi(-x,y,z) = \varphi(x,y,z)$, and

$$\langle x\bar\varphi(x,y,\bar z)\rangle \equiv 0.$$

Hence, the relative strength of neglected terms would be ~ $10^{-3}$.

[10] With averaging over the phases of the betatron oscillations defined as $(2\pi)^2\langle f\rangle_{\phi_x,\phi_y} \equiv \int_0^{2\pi}\int_0^{2\pi} f d\phi_x d\phi_y$

Averaging (A12) over betatron phases is straightforward using a well-known expression:

$$F(\xi) = \left\langle e^{-2\xi\cos^2\psi} \right\rangle_\psi \equiv \frac{1}{2\pi}\int_0^{2\pi} d\psi\, e^{-2\xi\cos^2\psi} = e^{-\xi} \cdot I_o(\xi) \quad (A16)$$

where $I_o(\xi)$ is the modified Bessel function of first kind [73]. Expressing the beam sizes through the lattice functions $\beta_{x,y}$ and geometric emittances $\varepsilon$[11], transverse dispersion $D$ and the RMS energy spread $\sigma_\delta$:

$$\sigma_x^2(s) = \beta_x(s)\varepsilon_x + D_x^2(s)\sigma_\delta^2; \quad \sigma_y^2(s) = \beta_y(s)\varepsilon_y; \quad (A17)$$

and using eq. (A16) we get:

$$\left\langle \Delta\tilde{h}_{sc} \right\rangle_{\varphi_{x,y}} = \frac{1}{\beta_o^2\gamma_o^3}\frac{Z^2 N_o r_c}{\sqrt{2\pi}\sigma_z} e^{-\frac{(s-\beta_o\tau)^2}{2\sigma_z^2}} \cdot \int_0^\infty d\vartheta \frac{\left\langle \exp\left(-\frac{2\beta_x I_x \cos^2\varphi_x}{2(1+\vartheta/u)\sigma_x^2}\right)\right\rangle_{\varphi_x} \left\langle \exp\left(-\frac{2\beta_y I_y \cos^2\varphi_y}{2(1+\vartheta/u)\sigma_y^2}\right)\right\rangle_{\varphi_y}}{\sqrt{(1+\vartheta/u)(1+\vartheta u)}}$$

and

$$\left\langle \Delta h_{sc} \right\rangle_{\phi_x,\phi_y} = \frac{1}{\beta_o^2\gamma_o^3}\frac{Z^2 N_o r_p}{A\sqrt{2\pi}\sigma_z} e^{-\frac{(s-\beta_o\tau)^2}{2\sigma_z^2}} \int_0^\infty d\vartheta \frac{F\left(\frac{A_x^2 \cdot f_E}{2(1+\vartheta/u)}\right) F\left(\frac{A_y^2}{2(1+\vartheta u)}\right)}{\sqrt{(1+\vartheta/u)(1+\vartheta u)}}; \quad (A18)$$

where we introduced new parameters:

$$A_{x,y}^2 \equiv \frac{I_{x,y}}{\varepsilon_{x,y}}; \quad f_E(s) = \frac{\beta_x(s)\varepsilon_x}{\beta_x(s)\varepsilon_x + D_x^2(s)\sigma_\delta^2} \equiv \frac{\beta_x(s)I_x}{\sigma_x^2 A_x^2} \quad (A19)$$

As seen from eq. (A18), the tune shifts depend on the horizontal and vertical actions as well as on the longitudinal position $s - \beta_o\tau$ inside the bunch. Averaging eq. (A13) over the betatron phases, one gets the Hamiltonian of the linear motion:

$$\Delta\tilde{h}_{sc} = -\frac{1}{\beta_o^2\gamma_o^3}\frac{Z^2 N_o r_p}{A\sqrt{2\pi}\sigma_z} e^{-\frac{(s-\beta_o\tau)^2}{2\sigma_z^2}} \left( \frac{2\beta_x(s)I_x \left\langle \cos^2(\psi_x(s)+\phi_x) \right\rangle}{\sigma_x(\sigma_x+\sigma_y)} + \frac{\left\langle y^2 \right\rangle}{\sigma_y(\sigma_x+\sigma_y)} \right)$$

and

$$\left\langle \Delta h_{scL} \right\rangle = -\frac{1}{\beta_o^2\gamma_o^3}\frac{Z^2 N_o r_p}{A\sqrt{2\pi}\sigma_z} e^{-\frac{(s-\beta_o\tau)^2}{2\sigma_z^2}} \frac{1}{\sigma_x(s)+\sigma_y(s)}\left( \frac{\beta_x(s)I_x}{\sigma_x(s)} + \frac{\beta_y(s)I_y}{\sigma_y(s)} \right). \quad (A20)$$

We define the average of a periodic function [12] $g(s) = g(s+C)$ as:

---

[11] With normalized emittances defined as $\varepsilon_n \equiv \gamma_o\beta_o\varepsilon$.

$$\langle g \rangle_C \equiv \frac{1}{C} \int_o^C g(s)\,ds. \tag{A21}$$

Using (A15) and (A20) we obtain the tune shifts experienced by particles near the axis then given by:

$$\delta Q_x = \frac{C}{2\pi}\left\langle \frac{\partial}{\partial I_x}\langle \Delta h_{sc}\rangle \right\rangle_C = -\frac{C}{2\pi}\frac{1}{\beta_o^2\gamma_o^3}\frac{Z^2 N_o r_p}{A\sqrt{2\pi}\sigma_z} e^{-\frac{(s-\beta_o\tau)^2}{2\sigma_z^2}} \left\langle \frac{1}{\sigma_x(s)+\sigma_y(s)}\frac{\beta_x(s)}{\sigma_x(s)} \right\rangle_C$$

$$= -\frac{C}{4\pi}\frac{1}{\varepsilon_x\beta_o^2\gamma_o^3}\frac{Z^2 N_o r_p}{A\sqrt{2\pi}\sigma_z} e^{-\frac{(s-\beta_o\tau)^2}{2\sigma_z^2}} \left\langle \frac{2 f_E(s)}{1+\frac{1}{u}} \right\rangle_C$$

to get

$$\delta Q_x = -Q_{sc}^x \cdot e^{-\frac{(s-\beta_o\tau)^2}{2\sigma_z^2}} \left\langle \frac{2\beta_x(s)\varepsilon_x}{\sigma_x(s)(\sigma_x(s)+\sigma_y(s))} \right\rangle_C = -Q_{sc}^x \cdot e^{-\frac{(s-\beta_o\tau)^2}{2\sigma_z^2}} \left\langle \frac{2 f_E(s) u(s)}{1+u(s)} \right\rangle_C$$

$$\delta Q_y = -Q_{sc}^y \cdot e^{-\frac{(s-\beta_o\tau)^2}{2\sigma_z^2}} \left\langle \frac{2\beta_y(s)\varepsilon_y}{\sigma_y(s)(\sigma_x(s)+\sigma_y(s))} \right\rangle_C = -Q_{sc}^y \cdot e^{-\frac{(s-\beta_o\tau)^2}{2\sigma_z^2}} \left\langle \frac{2}{1+u(s)} \right\rangle_C \tag{A22}$$

where we introduce an approximate values for space-charge-induced tune spreads as:

$$Q_{sc}^{x,y} = \frac{C}{4\pi\beta_o^2\gamma_o^3\varepsilon_{x,y}}\frac{Z^2 N_o r_p}{A\sqrt{2\pi}\sigma_z}. \tag{A23}$$

Similarly, using (A15) and (A18-A19) we obtain the tune shifts for arbitrary amplitudes of betatron oscillations:

$$\Delta Q_{scx} = Q_{sc}^x\, e^{-\frac{(s-\beta_o\tau)^2}{2\sigma_z^2}} \cdot \int_0^\infty d\vartheta \left\langle f_E \frac{F'\left(\frac{A_x^2 \cdot f_E}{2(1+\vartheta/u)}\right) F\left(\frac{A_y^2}{2(1+\vartheta u)}\right)}{\sqrt{(1+\vartheta/u)^3(1+\vartheta u)}} \right\rangle_C;$$

$$\Delta Q_{scy} = Q_{sc}^y\, e^{-\frac{(s-\beta_o\tau)^2}{2\sigma_z^2}} \cdot \int_0^\infty d\vartheta \left\langle \frac{F\left(\frac{A_x^2 \cdot f_E}{2(1+\vartheta/u)}\right) F'\left(\frac{A_y^2}{2(1+\vartheta u)}\right)}{\sqrt{(1+\vartheta/u)(1+\vartheta u)^3}} \right\rangle_C; \tag{A24}$$

---

[12] Note that lattice functions $\beta_{x,y}(s+C) = \beta_{x,y}(s)$ are periodic function of the ring circumference, $\psi_{x,y}(s+C) = \psi_{x,y}(s) + 2\pi Q_{x,y}$ while betatron phases are monotonically growing functions.

with $F'(\xi) = \dfrac{dF(\xi)}{d\xi} = e^{-\xi}(I_1(\xi) - I_o(\xi))$. Plot in Fig. A1 shows $F(A^2)$ and $-F'(A^2)$ as functions of $A$ (e.g. for $q=0$, $A$ is in the units the RMS beam size). The behavior of the kernel in integral (A24), $F'F$, indicate that the tune shifts diminish for particles with large oscillating amplitudes, e.g. $\Delta Q_{sc\,x} -> 0$ when $I_{x,y} \to \infty$. It also indicates that the tune spread for particles confined within $A_{x,y} \leq 5$ is close to that of the maximum shown in eq. (A22).

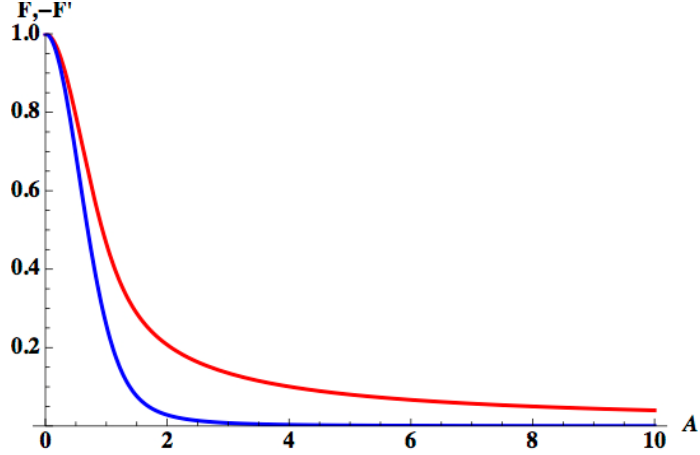

Fig. A1. Plots of $F(A^2)$ (red) and $-F'(A^2)$ (blue) as function of $A$ (horizontal axis).

For further use in the paper we re-write expressions for the tune shifts as:

$$\Delta Q_{sc\,x,y} = -Q_{sc}^{x,y} \cdot e^{-\dfrac{(s-\beta_o \tau)^2}{2\sigma_z^2}} \cdot f\!f_{x,y}\left(\dfrac{I_x}{\varepsilon_x}, \dfrac{I_y}{\varepsilon_y}\right);$$

$$f\!f_x(A_x^2, A_y^2) = -\int_0^\infty d\vartheta \left\langle f_E \dfrac{F'\left(\dfrac{A_x^2 \cdot f_E}{2(1+\vartheta/u)}\right) F\left(\dfrac{A_y^2}{2(1+\vartheta u)}\right)}{\sqrt{(1+\vartheta/u)^3(1+\vartheta u)}} \right\rangle_C ; \qquad (A25)$$

$$f\!f_y(A_x^2, A_y^2) = -\int_0^\infty d\vartheta \left\langle \dfrac{F\left(\dfrac{A_x^2 \cdot f_E}{2(1+\vartheta/u)}\right) F'\left(\dfrac{A_y^2}{2(1+\vartheta u)}\right)}{\sqrt{(1+\vartheta/u)(1+\vartheta u)^3}} \right\rangle_C .$$

We want to note that both $f\!f_x$ and $f\!f_y$, while being a smooth functions of $I_{x,y}$, are neither a simple exponential or can be expressed as a well-known functions. It means that in practice they should be calculated (tabulated) for a specific storage ring.

Finally, let us note that if the contribution of the energy spread term in (A17) is negligible, i.e.

$$\sigma \beta_{x,y}(s)\varepsilon_{x,y} \gg D^2_{x,y}(s)\sigma^2_\delta,$$

then both (A24) and (A25) can be reduced to a "symmetrically looking" expressions with $f_E = 1$. Further, for a round beam with equal transverse emittances $\varepsilon_x = \varepsilon_y = \varepsilon$ we can write eq. (A22) as:

$$\delta Q_x = -Q_{sc}\, e^{-\frac{(s-\beta_o\tau)^2}{2\sigma_z^2}} \left\langle 2\left(1+\sqrt{\frac{\beta_y(s)}{\beta_x(s)}}\right)^{-1}\right\rangle_C$$

$$\delta Q_y = -Q_{sc}\, e^{-\frac{(s-\beta_o\tau)^2}{2\sigma_z^2}} \left\langle 2\left(1+\sqrt{\frac{\beta_x(s)}{\beta_y(s)}}\right)^{-1}\right\rangle_C.$$

(A26)

with expressions

$$Q_{sc} = \frac{C}{4\pi\beta_o^2\gamma_o^3\varepsilon}\frac{Z^2 N_o r_p}{A\sqrt{2\pi}\sigma_z};$$

(A28)

frequently used as an estimate for tune spread induced by space charge in hadron storage rings.

To connect the particle's distribution with the beam current, we should note that

$$I(s,\tau) = \hat{I}\exp\left(-\frac{(s-\beta_o\tau)^2}{2\sigma_z^2}\right) \equiv \hat{I}\exp\left(-\frac{(s/v_o - t)^2}{2\sigma_t^2}\right);$$

$$\hat{I} \equiv \frac{eZN_o v_o}{(2\pi)^{1/2}\sigma_z};\quad \sigma_t = \frac{\sigma_z}{v_o};$$

(A29)

$$f(x,y,s,\tau=ct) = \frac{\hat{I}}{2\pi\sigma_x\sigma_y Z e v_o}\exp\left(-\frac{x^2}{2\sigma_x^2} - \frac{y^2}{2\sigma_y^2} - \frac{(s-\beta_o\tau)^2}{2\sigma_z^2}\right);$$

where $\hat{I}$ is the beam's peak current. This allows rewriting eq. (A23) as

$$Q_{sc}^{x,y} = \frac{C}{4\pi\beta_o^3\gamma_o^2\varepsilon_{x,y}}\frac{Z}{A}\frac{\hat{I}}{I_{pA}};\quad I_{pA} = \frac{m_p c^3}{e} \equiv \frac{ec}{r_p} \cong 31.3\cdot 10^6\ \text{A},$$

(A30)

where $I_{pA}$ is Alfven current redefined for protons.

**Appendix B. Transverse distribution of the electron beam**

We are proposing using a low energy electron beam to compensate the space charge tune shift of hadron beam with much higher energy with a similar beam size and peak currents. Hence, the focusing effect of the hadron beam on the electron beam will be much stronger and, most likely, will lead to pinching of the electron beam. The result of such interaction will be neither desirable nor controllable. There is a practical way of keeping electron beam transverse distribution controllable and fixed [26] – the propagating of magnetized beam in a strong solenoid field [77-78]. In this case, with an appropriately chosen strength of the solenoid, the electron beam profile can be both maintained and controlled.

Similarly to the arguments used for calculating the EM field induced by the hadron beam, we can conclude that the 4-potential induced by the electron beam is given by (A5) with the change of the kinematic variables: $\gamma_o \to \gamma_e; \beta_o \to \beta_e$

$$\left(\varphi_e(x,y,z,\tau), \vec{A}_e(x,y,z,\tau)\right)_{lab} = \gamma_e(1, \beta_e \hat{s}) \cdot \bar{\varphi}_e(x, y, \gamma_e(z - \beta_e \tau)) \tag{B1}$$

with the scalar potential $\bar{\varphi}_e$ satisfying the reduced Poison equation in the e-beam co-moving frame:

$$\Delta_\perp \bar{\varphi}_e(\vec{r}_\perp, \bar{z}) = -4\pi \rho_e(\vec{r}_\perp, \bar{z}) . \tag{B2}$$

As we discussed in the paper, the convolution of the longitudinal profile of the electron bunch has to fit that of the hadron bunch, while the transverse profile should be the same, e.g.

$$\rho_e(\vec{r}_\perp, \bar{z}) = \rho_{e\perp}(\vec{r}_\perp) \cdot \bar{g}(\bar{z}); \quad \bar{\varphi}_e(\vec{r}_\perp, \bar{z}) = \bar{\varphi}_{e\perp}(\vec{r}_\perp) \cdot \bar{g}(\bar{z});$$
$$\gamma_e \bar{g}(\bar{z}) = g(z - \beta_e \tau); \quad \Delta_\perp \bar{\varphi}_{e\perp}(\vec{r}_\perp) = -4\pi \rho_{e\perp}(\vec{r}_\perp). \tag{B3}$$

To satisfy the requirement for a desirable $\bar{\varphi}_e(\vec{r}_\perp)$ one has to "simply" provide a beam with transverse density distribution satisfying

$$\rho_{e\perp}(\vec{r}_\perp) = -\frac{\Delta_\perp \bar{\varphi}_{e\perp}(\vec{r}_\perp)}{4\pi}. \tag{B4}$$

It is easy to see that negatively charged beam is needed to compensate for the space charge of a positively charged particles, e.g. we know the sign of $\rho_e < 0$. But while looking as a simple mathematics (i.e. double differentiation of a given function), one should take into account that the density should not change the sign, i.e. on potentials with (unless the ignored longitudinal part of the Poisson equation become non-negligible for some reasons)

$$\Delta_\perp \bar{\varphi}_e(\vec{r}_\perp) \geq 0; \quad \forall \vec{r}_\perp$$

are allowed in practice. Unfortunately, this is not the only limitation – generating an arbitrary profile of the electron bunch is a non-trivial engineering undertaking.

While there are practical challenges of generating the desirable transverse profiles (especially when the current is modulated) a large diversity of transverse profiles had been generated in practice [77-25]. Hence, we can assume a smooth profile mimicking to a significant degree the distribution of a hadron beam can be generated.

It most likely means that only the main features of the space charge induced tune spread can be mimicked by the compensating beam and only partial compensation of the tune spread would be possible. The degree of compensation will depend both on the ring and the attainable electron beam profile.

**Appendix C. Tune shift induced by the e-beam compensator**

Let's consider an electron beam in a compensating scheme with a known 4-potential (see (B1)):

$$\left(\varphi_e(x,y,s,\tau), \vec{A}_e(x,y,s,\tau)\right)_{lab} = \gamma_e (1, \beta_e \hat{s}) \cdot \bar{\varphi}_e \left(x, y, \gamma_e (s - \beta_e \tau)\right) \quad (C1)$$

and according (B2)

$$\left(\varphi_e(x,y,s,\tau), \vec{A}_e(x,y,s,\tau)\right)_{lab} = \gamma_e (1, \beta_e \hat{s}) \cdot \bar{\varphi}_{e\perp}(x, y) \cdot g(s - \beta_e \tau);$$

$$\int_{-\infty}^{\infty} g(s) ds = 1. \quad (C2)$$

The later can be input into the transverse accelerator Hamiltonian for the hadrons (similarly to the procedure in Appendix A) to find how it changes:

$$\delta \tilde{h}_e = \frac{Ze\gamma_e}{\beta_0 p_o c} (1 - \beta_e \beta_0) \bar{\varphi}_{e\perp}(x, y) \cdot g(s - \beta_e \tau). \quad (C2)$$

Assuming the symmetry of the electron beam distribution, we can assume that

$$\bar{\varphi}_e(x, y) = \bar{\varphi}(x^2, y^2) \quad (C4)$$

and we can average the Hamiltonian over the betatron phases to find the effective Hamiltonian:

$$\langle \delta h_e \rangle = \frac{Z\gamma_e e}{\beta_0 p_o c} (1 - \beta_e \beta_0) \left\langle \bar{\varphi}\left(2I_x \beta_x(s)\cos^2\phi_x, 2I_y \beta_y(s)\cos^2\phi_y\right)\right\rangle_{\phi_{x,y}} \cdot g(s - \beta_e \tau). \quad (C5)$$

The effective Hamiltonian for a hadron traversing the interaction region:

$$s = \beta_o \cdot (\tau - \tau_o); \quad \tau = \tau_o + \frac{s}{\beta_o};$$

$$\int_{-L/2}^{L/2} ds \langle \delta h_e \rangle = \frac{Ze\gamma_e}{\beta_0^2 \gamma_o Amc^2}(1-\beta_e\beta_0) \times$$
$$\int_{-L/2}^{L/2} ds \langle \overline{\varphi}(2I_x\beta_x(s)\cos^2\phi_x, 2I_y\beta_y(s)\cos^2\phi_y) \rangle_{\phi_{x,y}} \cdot g\left(s\left(1-\frac{\beta_e}{\beta_0}\right) - \beta_e\tau_o\right) \quad \text{(C5)}$$

Further evaluation is impossible without specific transverse distribution of electron beam. It is natural to assume that it is also Gaussian,

$$f_e(x,y,s,\tau=ct) = \frac{N_e}{(2\pi)^{3/2}\sigma_x\sigma_y}\exp\left(-\frac{x^2}{2\sigma_x^2}-\frac{y^2}{2\sigma_y^2}\right)g(s-\beta_e\tau); \quad \text{(C6)}$$

$$\varphi(\vec{r}) = -eN_e g(s-\beta_e\tau)\int_0^\infty \frac{e^{-\frac{x^2}{2\sigma_{ex}^2+q}-\frac{y^2}{2\sigma_{ey}^2+q}}}{\sqrt{(2\sigma_{ex}^2+q)(2\sigma_{ey}^2+q)}} dq.$$

and continue with the Hamiltonian variation of

$$\langle \Delta h_e \rangle = \frac{ZN_e r_p}{\beta_0^2 \gamma_o A}(1-\beta_e\beta_0) \cdot g\left(s\left(1-\frac{\beta_e}{\beta_0}\right)-\beta_e\tau_o\right) \times \int_0^\infty dq \frac{F\left(\frac{\beta_x(s)I_x}{q+2\sigma_x^2}\right)F\left(\frac{\beta_y(s)I_y}{q+2\sigma_y^2}\right)}{\sqrt{(q+2\sigma_x^2)(q+2\sigma_y^2)}}; \quad \text{(C7)}$$
$$F(x) = I_o(x)e^{-x};$$

and the induced integral Hamiltonian is

$$\Delta H = \int_{-L/2}^{L/2} \langle \Delta h_e \rangle ds = \frac{ZN_e r_p}{\beta_0^2 \gamma_o A}(1-\beta_e\beta_0) \times$$
$$\int_{-L/2}^{L/2} ds\, g\left(s\left(1-\frac{\beta_e}{\beta_0}\right)-\beta_e\tau_o\right)\int_0^\infty dq \frac{F\left(\frac{\beta_x(s)I_x}{q+2\sigma_x^2}\right)F\left(\frac{\beta_y(s)I_y}{q+2\sigma_y^2}\right)}{\sqrt{(q+2\sigma_x^2)(q+2\sigma_y^2)}}; \quad \text{(C8)}$$

and the induced tune shifts are given by following convolutions:

$$\Delta Q_x = \frac{ZN_e r_p}{2\pi\beta_0^2 \gamma_o A}(1-\beta_e\beta_0) \int_{-L/2}^{L/2} ds g(\xi) \int_0^\infty dq \beta_x \frac{F'\left(\frac{\beta_x I_x}{q+2\sigma_x^2}\right) F\left(\frac{\beta_y I_y}{q+2\sigma_y^2}\right)}{\sqrt{(q+2\sigma_x^2)^3(q+2\sigma_y^2)}};$$

$$\Delta Q_x = \frac{ZN_e r_p}{2\pi\beta_0^2 \gamma_o A}(1-\beta_e\beta_0) \int_{-L/2}^{L/2} ds g(\xi) \int_0^\infty dq \beta_y \frac{F\left(\frac{\beta_x I_x}{q+2\sigma_x^2}\right) F'\left(\frac{\beta_y I_y}{q+2\sigma_y^2}\right)}{\sqrt{(q+2\sigma_x^2)(q+2\sigma_y^2)^3}};$$

$$\xi = s\left(1-\frac{\beta_e}{\beta_0}\right) - \beta_e\tau_o.$$

(C9)

One of possible simplification to (C9) can be done for a case when the electron beam size stays constant through the interaction region and $\beta_{x,y} \gg L$, e.g. one can use a thin lens approximation for the kick. In this approximation the effect of the slippage and the effect of the kick are separable:

$$\Delta Q_x \cong \frac{ZN_e r_p \bar{\beta}_x}{2\pi\beta_0^2 \gamma_o A}(1-\beta_e\beta_0)\, G(\tau) \int_0^\infty dq \frac{F'\left(\frac{\bar{\beta}_x I_x}{q+2\sigma_x^2}\right) F\left(\frac{\bar{\beta}_y I_y}{q+2\sigma_y^2}\right)}{\sqrt{(q+2\sigma_x^2)^3(q+2\sigma_y^2)}};$$

$$\Delta Q_y \cong \frac{ZN_e r_p \bar{\beta}_y}{2\pi\beta_0^2 \gamma_o A}(1-\beta_e\beta_0)\, G(\tau) \int_0^\infty dq \frac{F\left(\frac{\bar{\beta}_x I_x}{q+2\sigma_x^2}\right) F'\left(\frac{\bar{\beta}_y I_y}{q+2\sigma_y^2}\right)}{\sqrt{(q+2\sigma_x^2)(q+2\sigma_y^2)^3}}; \quad (C10)$$

$$G(\tau) = \int_{-L/2}^{L/2} ds\, g\left(\left(1-\frac{\beta_e}{\beta_0}\right)s - \beta_e\tau\right); \quad \bar{\beta}_{x,y} = \frac{1}{L}\int_{-L/2}^{L/2} \beta_{x,y}(s)ds.$$

The idea of the method is that $G(\tau)$ approximates the longitudinal shape of the hadron beam. Than the goal of the transverse shaping and choosing appropriate $\beta_{x,y}$ and electron beam intensity and the sizes is to approximate the tune shift values induced by the space charge effects and their dependence on the transverse actions. These assumptions set already familiar requirements:

$$G(\tau) = e^{-\frac{(s-\beta_o\tau)^2}{2\sigma_z^2}}; N_e = \frac{ZN_o}{\gamma_o^2(1-\beta_e\beta_0)} \frac{C}{\sqrt{2\pi}\sigma_z}. \quad (C11)$$

**Appendix D. Deconvolution.**

Before starting the derivation for deconvolving eq. (22), let us list our assumptions:

1. By definition, $\Delta z = c\Delta t > 0$;

2. We assume that the hadron beam longitudinal distribution function, $q(z)$, is an analytical function, with finite integral and values diminishing at infinity;
3. Furthermore, we assume that at large distances its derivative also vanishes faster that $1/|z|$, e.g. that $|q'(z)| < A/|z|^{1+\varepsilon}$, $\varepsilon > 0$. This rather weak assumption will be used for proving the convergence of the convolution.

To de-convolve the equation (22), let's rewrite it in following form:

$$\int_0^{\Delta z} g(z+\zeta)d\zeta = q(z)$$
$$\Downarrow \qquad (D1)$$
$$\int_z^{z+\Delta z} g(\zeta)d\zeta = q(z);$$

Taking derivative we get finite step differential equation on $g$:

$$g(z+\Delta z) - g(z) = q'(z), \qquad (D2)$$

which can be solved by turning the finite series

$$\sum_{m=0}^n q'(z+m\Delta z) = \sum_{m=0}^n g(z+(m+1)\Delta z) - \sum_{m=0}^n g(z+m\Delta z) = \sum_{m=1}^{n+1} g(z+m\Delta z) - \sum_{m=0}^n g(z+m\Delta z);$$
$$\sum_{m=0}^n q'(z+m\Delta z) = g(z+(n+1)\Delta z) - g(z) \qquad (D3)$$

assuming that $g(z)_{z\to\infty} \to 0$ we can derive the result:

$$g_+(z) = -\sum_{m=0}^\infty q'(z+m\Delta z), \qquad (D4)$$

We naturally assume that the sum $\sum_{m=0}^\infty q'(z+m\Delta z)$ converges, i.e. that the derivative to distribution function at large values falls faster than $A/|z|^{1+\varepsilon}$, $\varepsilon > 0$. Similarly, using

$$\sum_{m=1}^{n+1} q'(z-m\Delta z) = \sum_{m=0}^n g(z-m\Delta z) - \sum_{m=1}^{n+1} g(z-m\Delta z) = g(z) - g(z-(n+1)\Delta z), \quad (D5)$$

and assuming that $g(z)_{z\to-\infty} \to 0$ we can derive the second result:

$$g_-(z) = +\sum_{m=1}^\infty q'(z-m\Delta z). \qquad (D6)$$

Function $g_-$ in (D6) is not necessarily identical to $g_+$ in (D4) since the function $g(z)$ is not unique. Adding any periodical function with period $\Delta z$ and a zero integral:

$$g_1(z) = g(z) + p(z);$$
$$p(z+\Delta z) = p(z); \int_0^{\Delta z} p(z)dz = 0; \quad (D7)$$

does not change the property the is satisfy the convolution equation. But it definitely violates natural requirement that $g(z)_{z\to\pm\infty} \to 0$, e.g. the electron bunch length is finite.

By observing (D4) and (D6) one can conclude that for a properly behaving distribution function

$$|q'(z)| < A/|z|^{1+\varepsilon}, \quad \varepsilon > 0 \quad (D8)$$

$g_+(z)_{z\to+\infty} -> 0$ and $g_-(z)_{z\to-\infty} -> 0$. It is rather easy to prove considering $z > 0$:

$$|g_+(z>0)| < \sum_{m=0}^{\infty} |q'(z+m\Delta z)| < A\sum_{m=0}^{\infty} |z+m\Delta z|^{-(1+\varepsilon)} < A\sum_{m=M}^{\infty} |m\Delta z|^{-(1+\varepsilon)} = \frac{A}{\Delta z^{1+\varepsilon}} \sum_{m=M}^{\infty} \frac{1}{m^{1+\varepsilon}};$$

$$M = Floor\left[\frac{z}{\Delta z}\right] > \frac{z-\Delta z}{\Delta z};$$

$$\sum_{m=M}^{2M-1} \frac{1}{m^{1+\varepsilon}} < \sum_{m=M}^{2M-1} \frac{1}{M^{1+\varepsilon}} = \frac{1}{M^{\varepsilon}}; \sum_{m=2M}^{4M-1} \frac{1}{(2M)^{1+\varepsilon}} < \frac{1}{(2M)^{\varepsilon}} \to \sum_{m=M}^{\infty} \frac{1}{m^{1+\varepsilon}} \le \frac{1}{M^{\varepsilon}} \sum_{n=0}^{\infty} \frac{1}{(2^{\varepsilon})^n} = \frac{1}{M^{\varepsilon}} \frac{2^{\varepsilon}}{2^{\varepsilon}-1};$$

$$|g_+(z>0)| < \frac{A}{\Delta z^{1+\varepsilon}} \frac{1}{M^{\varepsilon}} \frac{2^{\varepsilon}}{2^{\varepsilon}-1} < \frac{A}{\Delta z^1} \frac{1}{(z-\Delta z)^{\varepsilon}} \frac{2^{\varepsilon}}{2^{\varepsilon}-1};$$

with the later expression definitely showing that $g_+(z)_{z\to+\infty} -> 0$. Proving $g_-(z)_{z\to-\infty} -> 0$ is identical in the logic with the given above prove, but for $z < 0$ replacing $M$ to $M = Floor\left[-\frac{z}{\Delta z}\right] > -\frac{z+\Delta z}{\Delta z}$.

While at some values of $\Delta z$ $g_-$ can be identical to $g_+$, there are clear physics examples of where they can diverge. For example, lets consider a bell-shaped $q(z)$ with its entire span falling within $\Delta z$, i.e. $q(\pm \Delta z/2) = 0$. In practice such distributions exist, for example a bunch in a single RF bucket. It means that

$$\int_{-\Delta z/2}^{\Delta z/2} q'(z)dz = 0; \quad q'\left(\pm\frac{\Delta z}{2}\right) = 0.$$

At the same time $q'(z)$ has at least one maximum and one minimum with an interval $\{-\Delta z/2, \Delta z/2\}$. Let's mark these $q'(z)$ extrema locations as

$z_+$ (maximum, $z_+ < 0$), $z_-$ (minimum, $z_- > 0$). Then for an arbitrary positive m we have:

$$g_+(z) = -\sum_{m=0}^{\infty} q'(z + m\Delta z), \, g_-(z) = +\sum_{m=1}^{\infty} q'(z - m\Delta z):$$

$$\begin{aligned} g_+(z_+ - m\Delta z) &= -q'(z_+) < 0 \\ g_+(z_- - m\Delta z) &= -q'(z_-) > 0 \\ g_-(z_+ + m\Delta z) &= q'(z_+) > 0 \\ g_-(z_- + m\Delta z) &= q'(z_-) < 0 \end{aligned} \quad , \tag{D9}$$

i.e. the oscillating nature of the $g_+$ at negative values and $g_-$ at positive values of argument. Furthermore,

$$\begin{aligned} g_+(z) &= 0; \, z > \Delta z / 2 \\ g_+(z) &= -q'(\Delta z \bmod(z/\Delta z + 1/2) - \Delta z / 2), \, z < \Delta z / 2 \\ g_-(z) &= 0; \, z < \Delta z / 2 \\ g_-(z) &= q'(\Delta z \bmod(z/\Delta z + 1/2) - \Delta z / 2), \, z > \Delta z / 2 \\ g_+(z) - g_-(z) &= -q'(\Delta z \bmod(z/\Delta z + 1/2) - \Delta z / 2) \end{aligned} \quad , \tag{D9}$$

with the difference being a periodic function with zero-value integral. The Fig. D1 below shows this behavior for a Gaussian $q(z)$ as function of the delay.

Since eq. (22) is linear, any combination of

$$g_\alpha(t) = \alpha g_+(t) + (1-\alpha) g_-(t)$$

is a deconvolution of the eq. (22). For practical application the mostly interesting is the an even sum of both,

$$2g_{1/2}(z) = g_+(z) + g_-(z) = \sum_{m=1}^{\infty} q'(z - m\Delta z) - \sum_{m=0}^{\infty} q'(z + m\Delta z)$$

$$2g_{1/2}\left(z + \frac{\Delta z}{2}\right) = \sum_{m=1}^{\infty} q'\left(z - m\Delta z + \frac{\Delta z}{2}\right) - \sum_{m=0}^{\infty} q'\left(z + m\Delta z + \frac{\Delta z}{2}\right)$$

$$\sum_{m=0}^{\infty} q'\left(z + m\Delta z + \frac{\Delta z}{2}\right) = \sum_{m=1}^{\infty} q'\left(z + m\Delta z - \frac{\Delta z}{2}\right)$$

$$2g_{1/2}\left(z + \frac{\Delta z}{2}\right) = \sum_{m=1}^{\infty} \left( q'\left(z - \left(m - \frac{1}{2}\right)\Delta z\right) - q'\left(z + \left(m - \frac{1}{2}\right)\Delta z\right) \right)$$

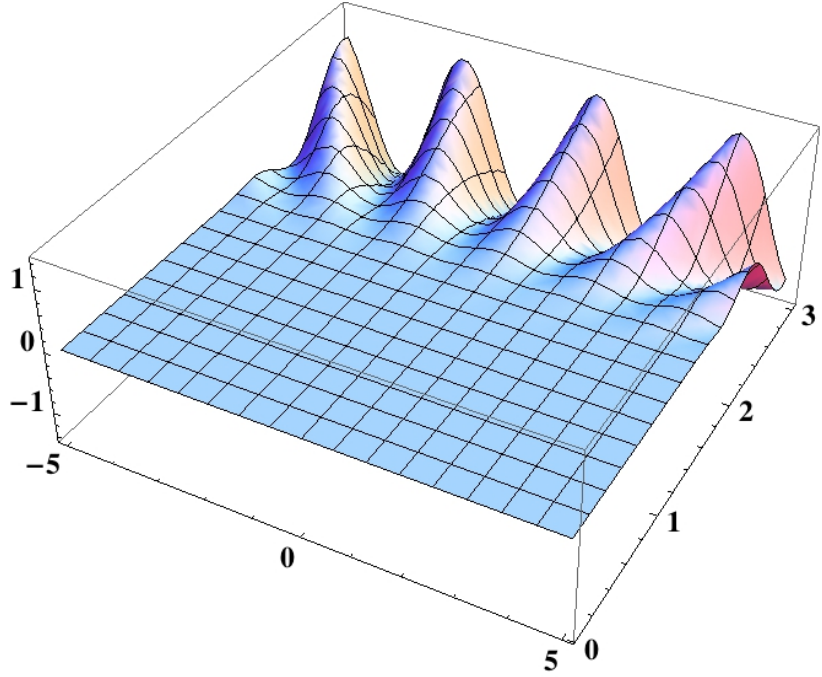

Fig. D1. 3D-plot of $\tau \cdot (g_+(t) - g_-(t))$ (vertical axis) for Gaussian convolution function with $t = z/\sigma_z \in (-5, 5)$ being a horizontal axis, and the third axis is $\tau = \Delta z/\sigma_z \in \{0, 3\}$ (see eqs. (D4), (D6)).

Hence, we can define a "symmetric-form" function as

$$G(z) = g\left(z + \frac{\Delta z}{2}\right) = \frac{1}{2}\sum_{m=1}^{\infty}\left(q'\left(z - \left(m - \frac{1}{2}\right)\Delta z\right) - q'\left(z + \left(m - \frac{1}{2}\right)\Delta z\right)\right) \quad (D10)$$

with easy prove

$$\sum_{m=1}^{\infty}\int_{z-\frac{\Delta z}{2}}^{z+\frac{\Delta z}{2}} dz_1 \left\{q'\left(z_1 - \left(m - \frac{1}{2}\right)\Delta z\right) - q'\left(z_1 + \left(m - \frac{1}{2}\right)\Delta z\right)\right\} = \sum_{m=1}^{\infty}\begin{Bmatrix} q(z + \Delta z - m\Delta z) - q(z + m\Delta z) \\ -q(z - m\Delta z) + q(z - \Delta z + m\Delta z) \end{Bmatrix}$$

$$= \left\{\sum_{m=0}^{\infty} q(z - m\Delta z) - \sum_{m=1}^{\infty} q(z + m\Delta z) - \sum_{m=1}^{\infty} q(z - m\Delta z) + \sum_{m=0}^{\infty} q(z + m\Delta z)\right\} = 2q(z)$$

that

$$\int_{z-\Delta z/2}^{z+\Delta z/2} G(\zeta) d\zeta = q(z) \,. \quad (D11)$$

If is $q(-z) = q(z)$ symmetric, then

$$q'(z) = -q'(-z) \to q'\left(z - \left(m - \frac{1}{2}\right)\Delta z\right) = -q'\left(\left(m - \frac{1}{2}\right)\Delta z - z\right)$$

$$q'\left(z + \left(m - \frac{1}{2}\right)\Delta z\right) \equiv q'\left(\left(m - \frac{1}{2}\right)\Delta z + z\right)$$

and $G(z)$ can be rewritten inform which is obviously symmetric terms:

$$G(z) = -\frac{1}{2}\sum_{m=1}^{\infty}\left(q'\left(\left(m - \frac{1}{2}\right)\Delta z - z\right) + q'\left(\left(m - \frac{1}{2}\right)\Delta z + z\right)\right) \tag{D12}$$

which is easy to prove:

$$v(z) = q'(a - z) + q'(a + z);$$
$$v(-z) = q'(a + z) + q'(a - z) \equiv v(z)$$

Let's consider a practical case with:

$$q(z) = \exp\left(-\frac{z^2}{2\sigma^2}\right); \tag{D13}$$

we have

$$g_+(z) = \frac{1}{\sigma^2}\sum_{n=0}^{\infty}(z + n\Delta z)\cdot\exp\left(-\frac{(z + n\Delta z)^2}{2\sigma^2}\right). \tag{D14}$$

It is obvious that the behavior of the $g(z)$ defined by $\tau = \frac{\Delta z}{\sigma}$:

$$g_+(t) = \frac{1}{\sigma}\sum_{n=0}^{\infty}(t + n\tau)\cdot\exp\left(-\frac{(t + n\tau)^2}{2}\right)$$
$$t = \frac{z}{\sigma}; \tau = \frac{\Delta z}{\sigma} \tag{D15}$$

Fig. D2 shows the value of $g_\pm(t)$ as functions of $t \in (-5,5)$ and $\tau \in \{0,3\}$. It is possible to see that for values of $\tau \in \{0,1\}$ the functions $g_\pm(t)$ remain positive. Fig. D3 shows details of $g_\pm(t)$ the $\tau = 1,2,3$. It is obvious that for $\tau > 2$ it is impossible to make $g(t) > 0$ approximating the required function. But it is also obvious that $\tau \leq 1.5$ eqs. (D4) and (D6) generate a smooth positive function, which can be, in principle, reproduced by a e-beam profile.

Figs. 3-5 in Chapter III also show that one can use $\tau \leq 1.5$ for generating smooth positive functions closely approximating the required forms. As shown in Fig. D4, the $g_+(t)$ for $\tau = 1.5$ get into a small negative values at $t < -3$. Detailed studies show that it becomes an oscillating sin-like function with amplitude about 1.5 $10^{-3}$ with period of $T \sim 1.5$. Simply cutting this tail at $t < -3$ makes a practically attainable positive function, whose convolution deviates from (22) by a small fraction $\sim 7\ 10^{-4}$.

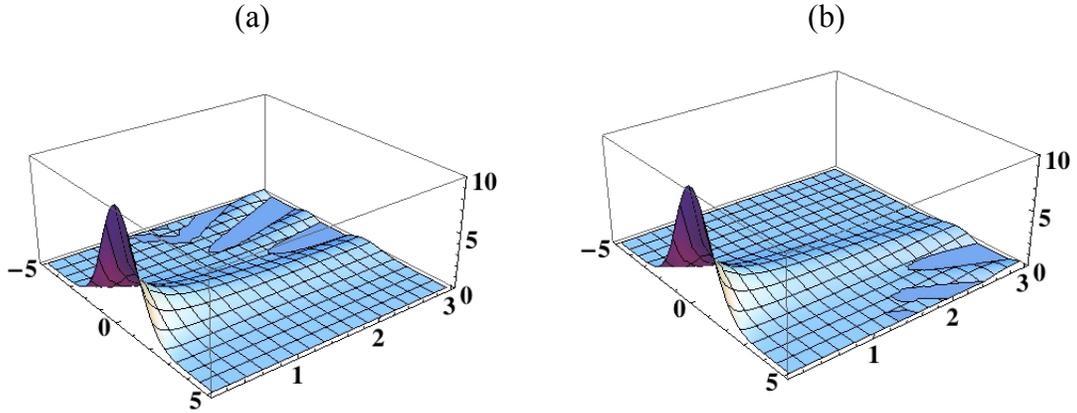

Fig. D2. 3D-plots of $g_+(t)$ (top) and $g_-(t)$ (bottom) as functions of $t \in (-5,5)$ and parameter $\tau \in \{0,3\}$. The clipping shows the area where function becomes negative.

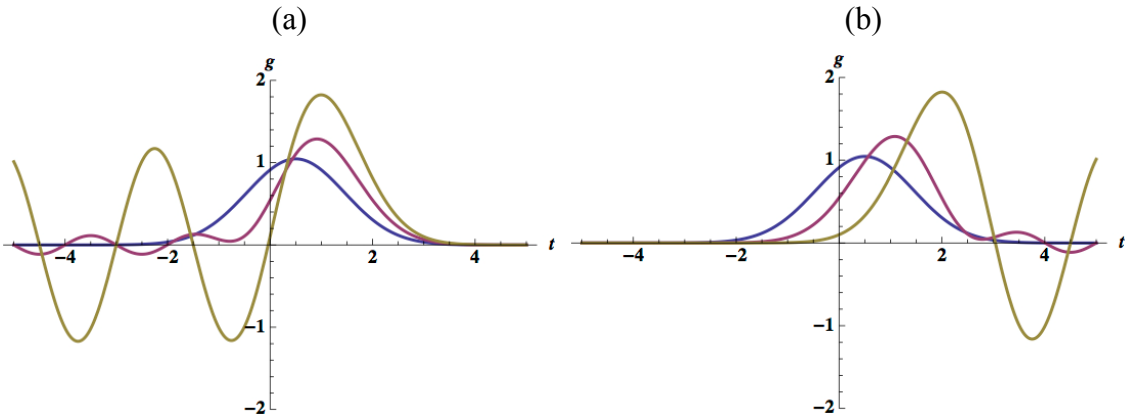

Fig. D3. Plot of $\tau \cdot g_\pm(t)$ ($g_+$ is on left and $g_-$ is on the right) as functions of $t \in (-5,5)$ and for $\tau = 1,2,3$.

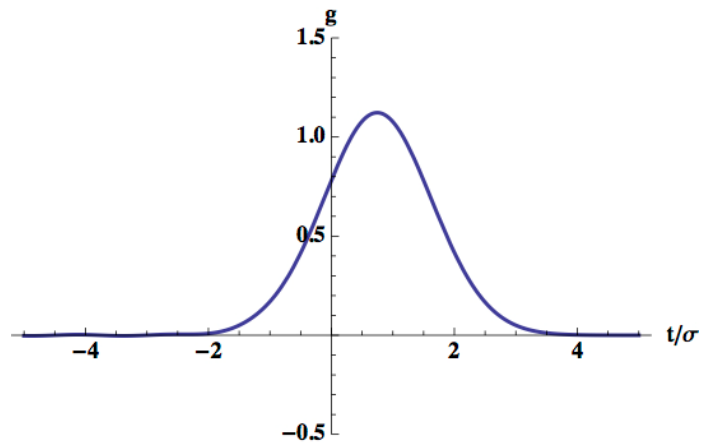

Fig. D4. 3D-plot of $\tau g_+(t)$ as function of $t \in (-5,5)$ and for $\tau = 1.5$.

As shown in Fig. D5, it is even possible to use $\tau = 2$ and smooth function to compensate more than 95% of the tune spread induced by the space charge. While this can be considered, using $\tau \leq 1.5$ is preferable for accurate space charge compensation.

(a) (b)

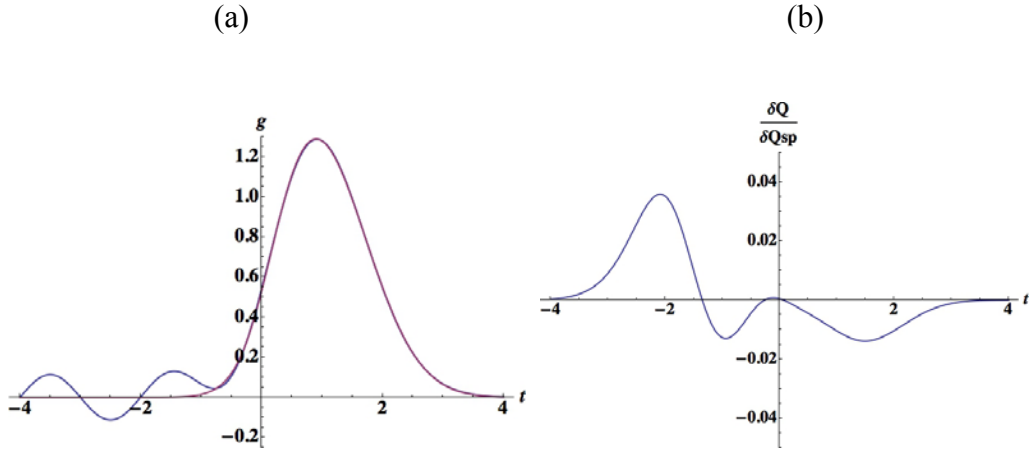

Fig. D5. (a) a plot of $g_+(t)$ for $\tau = 2$ (blue) and a fit of a positive function (magenta), (b) shows a residual tune shift as a function of the position within a hadron bunch, when a fitted function is used to compensate the space charge.